\documentclass{ws-mpla}

\newcommand{\eg}{e.g.~}
\newcommand{\lb}{\left(}
\newcommand{\rb}{\right)}

\newcommand{\eqn}{equation}
\newcommand{\al}{\alpha}
\newcommand{\be}{\beta}
\newcommand{\GeV}{{\ensuremath\,\rm GeV}\xspace}
\newcommand{\TeV}{{\ensuremath\,\rm TeV}\xspace}
\newcommand{\lam}{\lambda}
\newcommand{\fb}{{\ensuremath\,\rm fb}\xspace}

\usepackage{xspace}
\usepackage{multirow}

\usepackage[super]{cite}
\usepackage{xcolor}
\usepackage[verbose,hypertexnames=false]{hyperref}
\hypersetup{colorlinks=false,allbordercolors=blue,pdfborderstyle={/S/U/W 1}}

\begin{document}
\bibliographystyle{hunsrt}

\markboth{Tania Robens}{Interference effects in new physics searches}

\catchline{}{}{}{}{}
\title{Interference effects in new physics searches
}

  \author{Tania Robens\footnote{permanent address}}


  \address{
    Rudjer Boskovic Institute\\
    Bijenicka cesta 54\\
    10000 Zagreb, Croatia\\
trobens@irb.hr
}

\maketitle


\begin{abstract}
  Interference effects are an important consequence of a correct description in physics theories within and beyond the Standard Model (SM) of particle physics. However, many current theoretical descriptions as well as experimental searches neglect such effects, which can, among others, lead to an incorrect description of e.g. kinematical distributions, at least within the context of UV-complete models. In this review, I briefly discuss the current status and most common descriptions as well as existing studies of such effects, where I focus on models with extended scalar searches.\\
RBI-ThPhys-2026-02

\end{abstract}



\section{Introduction}

In new physics scenarios that include additional scalar bosons that can decay resonantly into SM like or new physics final states, very often the assumption is made that these resonances can be modelled separately and then added to the assumed SM background without taking the important interference between signal and background into account, which is a clear prediction of quantum mechanics and quantum field theory. Depending on the investigated final state, scenario, as well as selection cuts, such effects can play an important role and therefore cannot be a priori neglected, even under the assumption of narrow widths (see e.g. work in the context of SM Higgs production and decay \cite{Dixon:2003yb,Kauer:2012hd,Martin:2012xc,Martin:2013ula,Coradeschi:2015tna}).

In this short review, I showcase several examples that have been already provided in the literature that highlight the importance of interference effects in the search for heavy scalar resonances. I will demonstrate that in general such effects cannot be a priori neglected, but instead the importance needs to be assessed on a case by case study. For most scenarios, leading-order tools including such interference terms are readily available and should be used by the experimental collaborations for signal modelling.


\section{Short introduction to finite width and interference effects}
\label{sec:WidthInterference}

Remembering lectures from Quantum Field Theory (see e.g. \cite{Itzykson:1980rh,Peskin:1995ev}), from a quantum mechanical perspective, only stable particles can serve as the initial and final states of S-matrix elements. Such particles are described by wavefunctions for which $t\,\rightarrow\,\pm\,\infty$. The description of scattering processes with these particles in the inital and final state is therefore correct for such scenarios. However, it has become common to describe processes in a factorized approach, which is justified in case the narrow width approximation is justified, which we will discuss below. This means that a process

\begin{\eqn*}
  a\,b\,\rightarrow\,c\,d
\end{\eqn*}
that is also mediated via an s-channel resonance $S$ is described via

\begin{\eqn*}
a\,b\,\rightarrow\,S
  \end{\eqn*}
for the calculation of the production rate or
\begin{\eqn*}
  S\,\rightarrow\,c\,d
\end{\eqn*}
for the calculation of partial decay widths or branching ratios. The factorized approach is then given by
\begin{\eqn}\label{eq:fact}
  a\,b\,\rightarrow\,S\,\rightarrow\,c\,d.
  \end{\eqn}

There are several reason for using factorization in the description of a scattering process. One can e.g. be the high complexity of the final state involving integration in n-dimensional particle phase space. Another important and not yet resolved topic is the inclusion of higher-order contributions either for production or decay,  which might only be feasible in a factorized version. However, describing particle processes using eqn. (\ref{eq:fact}) is based on the assumption of on-shellness of particle $S$, and furthermore neglects additional contributions to the process
\begin{\eqn*}
  a\,b\,\rightarrow\,c\,d,
  \end{\eqn*}

or the interference of such contributions with the matrix element describing (\ref{eq:fact}) directly. Depending on the specific setup, the above assumptions lead to significant deviations from the correct description of the above process, in particular in variables that might be used to optimize cuts to distinguish signal from background.

\subsection{Narrow width approximation}
\label{sec:NWA}


The theoretical foundation for a factorized approach in describing a process as given in eqn. (\ref{eq:fact}) this is the so-called narrow width approximation (NWA) \cite{pilkuhn1967interactions,Dicus:1984fu}. We here briefly review the underlying assumptions (see also \cite{Feuerstake:2024uxs}).
For such a process, the matrix element can be written 
via the resummation of the infinite series of self-energy insertions $\Sigma(p^2)$ between the tree-level propagators $D(p^2)= i/(p^2 - m_S^2)$ (see Ref.~\cite{Peskin:1995ev,Denner:2019vbn})
as
\begin{\eqn}\label{eq:mabcd}
  \mathcal{M}_{a\,b\,\rightarrow\,c\,d}\,\,=\, \frac{1}{p^2 - m_{S,0}^2 + \hat{\Sigma}(p^2)}\,\mathcal{F}
  \sim\,\frac{1}{p^2-m_S^2+i\,\Gamma_S\,m_S}\,\mathcal{F}.
\end{\eqn}
In the above expression, $p$ denotes the four-momentum in the $s$-channel, and $\mathcal{F}$ depends on the four-momenta of the external particles and additional possible model parameters. Furthermore, $m_{S,0}, m_S$ are the tree-level mass and physical mass of the scalar, respectively. 

The optical theorem is then used to relate the quantum field theoretical renormalized self-energy, $\hat{\Sigma}$, to the physical total width of $S$, $\Gamma_S$, at the pole $p^2=m_S^2$:

\begin{\eqn}
  \text{Im} \hat{\Sigma}\,\lb p^2\,=\,m_S^2 \rb \,=\,\Gamma_S\,m_S.
  \end{\eqn}
In this expression, the total decay width $\Gamma$ signifies the sum over all possible partial decay widths\footnote{To ensure gauge invariance, it is in addition helpful to work in the complex mass scheme, where $m_S$ is taken to be complex, see \eg \cite{Denner:1999gp,Denner:2005fg,Denner:2006ic,Denner:2014zga,Denner:2019vbn} for details. 
}.

One important assumption in the narrow width approximation is that the total width of the particle is small compared to its mass. Mathematically, this corresponds to taking the limit of $\Gamma\,\rightarrow\,0$, giving
\begin{\eqn}\label{eq:NWA_delta}
  \frac{1}{|p^2-m^2+i\,m\,\Gamma|^2}\,\rightarrow\,\frac{\pi}{m\,\Gamma}\,\delta\lb p^2-m^2 \rb,
\end{\eqn}
leading to factorization and assuming the s-channel contribution to be on shell.

Further assumptions in the NWA are that the function $\mathcal{F}$ only obtains major contributions around $p^2\,\sim\,m^2$ 
and only varies mildly in the region $p^2\,\in\,\left[ \lb m-\Gamma\rb^2,\lb m+\Gamma \rb^2 \right]$\footnote{This assumption, and the validity of the NWA, can be violated \eg near kinematic thresholds.}. Combining these assumptions, we then obtain for the $s$-channel mediated cross section
\begin{align}
\sigma_{a\,b\,\rightarrow\,c\,d}^s\,\simeq\,\sigma_{a\,b\,\rightarrow\,S}\,\times\,\int^{p^2_\text{max}}_{p^2_\text{min}}\,\frac{d\,p^2}{2\,\pi}\,\frac{2\,m}{|p^2-m^2+i\,m\,\Gamma|^2}\,\times\,\Gamma_{S\,\rightarrow\,c\,d}.
\end{align}
Making use of Eq.~\eqref{eq:NWA_delta}, the limit $\Gamma\,\rightarrow\,0$ leads to the well-known factorisation into the on-shell production cross section times the branching ratio, 
\begin{\eqn}\label{eq:fac}
\sigma_{a\,b\,\rightarrow\,c\,d}^s\,\simeq\,\sigma_{a\,b\,\rightarrow\,S}\,\times\,\underbrace{\frac{\Gamma_{S\,\rightarrow\,c\,d}}{\Gamma}}_{\text{BR}\lb S\,\rightarrow\,c\,d \rb}.
\end{\eqn}
The formal error of the above expression is given by  $\mathcal{O}\lb \frac{\Gamma}{m} \rb$. However, several authors have highlightened the restrictions of the above assumptions, for example from off-shell and threshold effects, the impact of nearby resonances and non-factorisable contributions, see \eg Refs.~\cite{Berdine:2007uv,uhlemann_dipl,Uhlemann:2008pm,Cacciapaglia:2009ic, Fuchs:2015jwa,Fuchs:2017wkq,Bagnaschi:2018ofa,Hoang:2024oeq}. A more general discussion regarding the correct treatment of  unstable particles is provided in Ref.~\cite{Denner:2019vbn}. 

Furthermore, the result in Eq.~(\ref{eq:fac}) highlights the importance of using the physical value of the total width in order to correctly describe the underlying physics. 
In particular, using the total width as an input parameter $\Gamma_\text{in}$, \eg determined by the detector resolution, in turn leads to arbitrary branching ratios that can easily exceed 1 if $\Gamma\,\equiv\,\Gamma_\text{in}\,\leq\,\Gamma_{S\,\rightarrow\,c\,d}$, and therefore unphysical rates.
Instead, using a properly calculated total width as a prediction of the model parameters enables a correct physical description.   

\subsection{Interference effects}

The second related topic we want to discuss are interference effects. These are connected to but not necessarily coincide with the approximations made in the NWA. This means that although a process might be described correctly in the sense that all finite width effects are taken into account by using the full propagator, still additional processes not captured by the $s$-channel process and the interference with this channel might give important additional contributions. We want to reemphasize that a correct quantum-mechanical description always includes all intermediate final states that contribute to the respective process.

To introduce interference effects, we decompose the matrix element into the one containing the target process and possible additional contributions according to
\begin{\eqn}
\mathcal{M}_\text{tot}\,=\,\mathcal{M}_S\,+\,\mathcal{M}_\text{rest},
\end{\eqn}
where $\mathcal{M}_\text{rest}$ contains all additional diagrams.
The total squared matrix element is then given by
\begin{\eqn}
|\mathcal{M}_\text{tot}|^2\,=\,|\mathcal{M}_S|^2\,+\,|\mathcal{M}_\text{rest}|^2\,+\,\underbrace{2\,\text{Re}\left[ \mathcal{M}_S\,\mathcal{M}^*_\text{rest} \right]}_\text{Interference}.
\end{\eqn}

In case of dominance by an on-shell resonance $S$, the process is well-described by the first contribution, which can easily be extended in the case of several resonances. However, it is not a priori clear that this holds for all kinematic variables in the process, including differential distributions or specific regions of phase space. Therefore, in general, a priori all above contributions need to be taken into account. In particular, one has to carefully investigate the contributions of interference terms between the resonances in case several of these contribute as well as between the resonances and the non-resonant terms. The actual importance of these can only be determined on a case by case basis.

\section{Prominent models}
\subsection{Real singlet extension}

In this section, we focus on the most simple scenario where the scalar sector of the SM is extended by an additional field that transforms as a singlet under the SM gauge transformations. This model has been vastly discussed in the literature, see e.g. \cite{Pruna:2013bma,Robens:2015gla,Robens:2016xkb,Ilnicka:2018def,Feuerstake:2024uxs} for a version of the model where in addition to renormalizability an additional $\mathbb{Z}_2$ symmetry is imposed such that the number of free parameters after electroweak symmetry breaking is minimal. As the singlet transforms trivially under the SM gauge group, it does not participate in electroweak symmetry breaking. The free parameters of the model are then given by

\begin{\eqn*}  
v,\,v_s,\,m_h,\,m_H,\,\sin\al.
\end{\eqn*}
From the five free parameters above, the vacuum expectation value of the doublet, $v$, as well as one of the additional scalar masses are fixed by current measurements, leaving in total 3 free parameters, including the singlet vacuum expection value (vev) $v_s$ or a related quantity. In the above notation we assume a mass hierarchy $m_h\,\leq\,m_H$. The mixing angle $\sin\al$ describes the mixing between gauge and mass eigenstates.
Note that frequently the second vev is reparametrized in terms of an angle such that $\tan\be\,\equiv\,\frac{v}{v_s}$. Alternatively one can take the branching ratio of $H\,\rightarrow\,h\,h$ of the total or respective partial decay width as an input parameter (see e.g. \cite{Robens:2016xkb}). Our notation is such that $\sin\al\,=\,0$ corresponds to the SM decoupling scenario if $m_h\,\sim\,125\,\GeV$ is taken as the SM-like scalar. All couplings to gauge bosons and fermions are inherited from the SM-like doublet and rescaled by $\sin\,\al\,\lb \cos\,\al \rb$ for the heavy (light) scalar.

\subsection{Two Higgs doublet model}

Two Higgs Doublet Models (2HDMs) are another popular extension of the SM scalar sector.
In such models, the scalar sector is enhanced by a second complex scalar field that acts as a doublet under the SM gauge group. A typical notation is then given by
\begin{\eqn*}
\Phi_a\,=\,\lb \begin{array}{c}{\phi^+_a}\\{\lb v_a+ \phi_a\,+\,\imath\,\eta_a \rb/\sqrt{2}}  \end{array}\rb
\end{\eqn*}
where $a\,=\,1,2$, $v_a$ denotes the vacuum expectation value of the respective doublet, and the fields $\phi^+_a\,\lb \rho_a,\,\eta_a\rb$ are taken to be complex (real) respectively. After electroweak symmetry breaking, we are left with 5 physical scalar fields, denoted by
\begin{\eqn*}
\underbrace{h,\,H,}_{\text{neutral, CP even}} \underbrace{A}_{\text{neutral, CP odd}}, H^\pm,
\end{\eqn*}
where we already assumed that CP is conserved. The most generic renormalizable form of the model allow for both flavour changing neutral currents as well as CP violation. In order to forbid the former,  additional symmetries are imposed on the potential. We refer the reader to e.g. \cite{Branco:2011iw} and references therein for an overview on possible symmetry classes, as well as a general exhaustive overview on 2HDMs.

A typical benchmark in many experimental searches is the softly broken $\mathbb{Z}_2$ symmetric version. Here, the potential is given by
\begin{eqnarray}\label{eq:2hdmpot}
  V_\text{2HDM}& =&\ \mu_1^2 |\Phi_1|^2 + \mu_2^2 |\Phi_2|^2 - \mu_{12}^2\,\lb \Phi^\dagger_1\,\Phi_2+ \text{h.c.}\rb+ \frac{1}{2} \lambda_1 |\Phi_1|^4 + \frac{1}{2} \lambda_2 |\Phi_2|^4 \nonumber\\
&  +& \lambda_3 |\Phi_1|^2 |\Phi_2|^2 + \lambda_4 |\Phi_1^\dagger \Phi_2|^2 
    + \frac{1}{2} \lambda_5 \left[(\Phi_1^\dagger\Phi_2)^2+\text{h.c.}\right]\,. 
\end{eqnarray}
where all parameters are taken to be real. The potential has 8 free parameters; a typical choice is then given by $v\,\equiv\,\sqrt{v_1^2+v_2^2}\,\sim\,246\,\GeV$,  fixed from electroweak precision measurements. Also, as in the singlet scenario, either $h$ or $H$ has to mimic 
the 125 \GeV~ resonance measured by the LHC experiments. A standard choice for the remaining 6 free parameters is then
\begin{\eqn}\label{eq:2hdmpars}
m_{h/H},\,m_A,\,m_{H^\pm},\,\tan\,\be\,\equiv\,\frac{v_2}{v_1},\,\cos\lb \be-\al\rb,\,\mu_{12}^2.
\end{\eqn}
We here introduced the mixing angles $\be$ and $\al$ that are related to mixings of the gauge and mass eigenstates in the CP-odd and even neutral sectors, respectively\footnote{Note this is basis dependent; we refer the interested reader e.g. to \cite{Davidson:2005cw} for more details.}. 
It needs to be said that the mixing angle combination $\cos\lb \be-\al \rb$ is of particular relevance as the limits 0 and 1 correspond to the so-called alignment limit where one of the Higgs couples to electroweak gauge bosons as in the Standard Model. The actual value depends on the choice for the 125 \GeV resonance, $h_{125}$ at the LHC. The value of it is constrained to a region relatively close to 0 for $h\,\equiv\,h_{125}$ and close to 1 for $H\,\equiv\,h_{125}$.

The above potential does not yet specify the coupling of the scalar sector to fermions. There are four different Yukawa types that ensure flavour conservation. Type 1 denotes the scenario where all fermions couple to $\Phi_2$. In type 2, only up-type quarks couple to $\Phi_2$, while down-type quarks and lepton couple to $\Phi_1$. In lepton-specific scenarios, only leptons couple to $\Phi_1$, while quarks couple to $\Phi_2$. Finally, in flipped scenarios, only down-type quarks couple to $\Phi_2$, while up-type quarks and leptons couple to $\Phi_1$. Realization of these Yukawa types is guaranteed by enforcing specific transformation properties in the scalar and fermionic sectors under discrete $\mathbb{Z}_2$ symmetries, see e.g. \cite{Branco:2011iw} for details.
 A recent discussion on state of the art and constraints in such models can e.g. be found in \cite{Robens:2025nev}.

\subsection{Other extensions}
Obviously, many more extensions exist that can incorporate additional scalar bosons. In principle, any extension by a $SU(2)\,\otimes\,U(1)$ multiplet can lead to additional scalars states. However, with the number of additional terms equally the number free parameter rapidly increases. We therefore mainly concentrate on the models discussed above. Several use for example $\theta$ instead of $\alpha$ for the mixing angle in the singlet scenario.

\section{Specific final states}
In the following, we display several examples from the literature. Note that notation and conventions might differ from the ones introduced in the previous section. We added clarifying comments when necessary.
\subsection{Diboson final states: electroweak gauge bosons}
For the singlet extension, interference and finite width effects have been long known and discussed in the literature. One of the first examples studied in more detail was the decay of a heavy scalar into a pair of electroweak vector bosons, i.e. $W^+W^-$ or $Z\,Z$. This is of particular importance as the coupling of scalars to electroweak gauge bosons are related via an important sum rule \cite{Gunion:1990kf} stemming from unitarity requirements on the model.

\begin{figure}
\begin{center}
\begin{minipage}{0.45\textwidth}
\includegraphics[width=\textwidth]{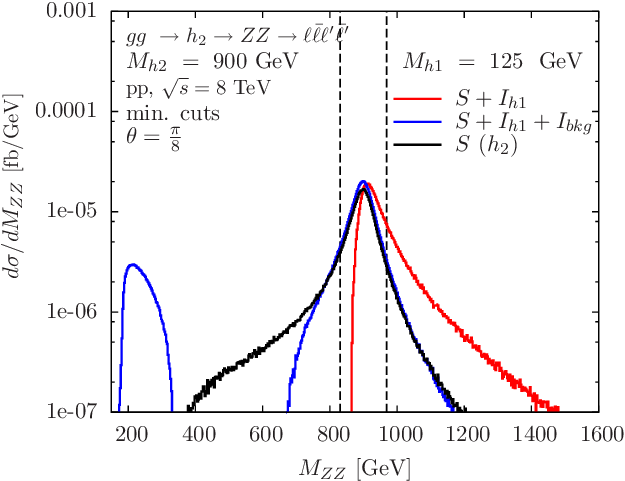}
\end{minipage}
\begin{minipage}{0.45\textwidth}
\includegraphics[width=\textwidth]{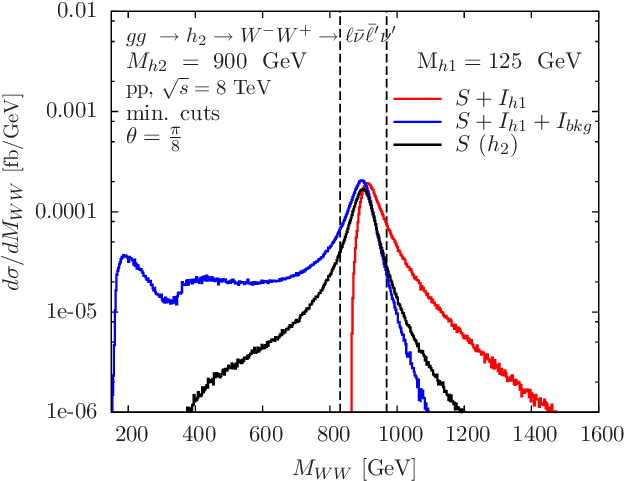}
\end{minipage}
\begin{minipage}{0.45\textwidth}
\includegraphics[width=\textwidth]{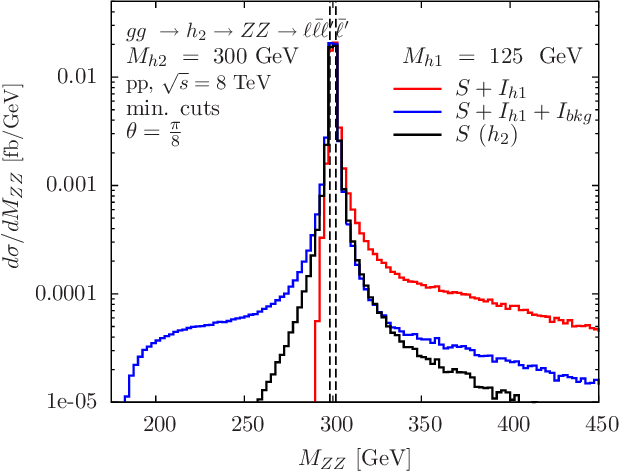}
\end{minipage}
\begin{minipage}{0.45\textwidth}
\includegraphics[width=\textwidth]{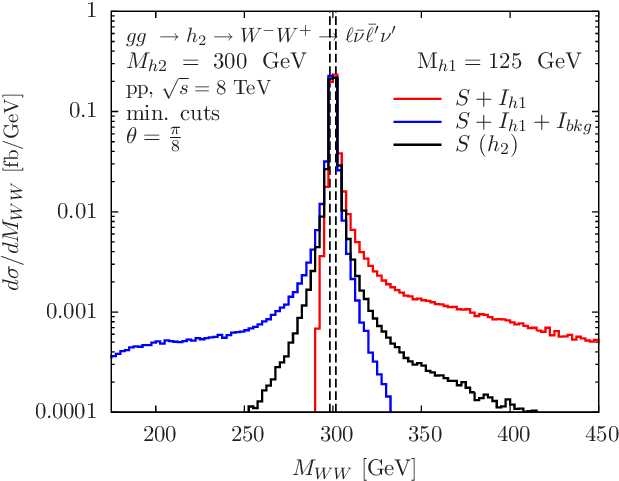}
\end{minipage}
\end{center}
\caption{\label{fig:mvvs} Invariant di-boson mass distributions for a scenario where $p\,p\,\rightarrow\,H\,\rightarrow\,V\,V$ is the target signature. In this work, $h1/2$ denotes the lighter/ heavier scalar state, and $\theta$ is the respective mixing angle. $S$ denotes the contribution from the target process, $I_{h1/bkg}$ are contributions from interference with the SM scalar, denote by $h1$, and the continuum background, respectively. Displayed is the signal-only distribution (in black), signal including interference from the 125 \GeV resonance (in red), as well as complete contribution where all interference terms are taken into account (in blue). The pure SM background is subtracted. Figure is taken from \cite{Kauer:2015hia}. See text and original reference for further details.} 
\end{figure}

We here chose to show results from \cite{Kauer:2015hia}, which discusses in detail the different contributions to the invariant di-boson mass distributions taking all possible contributions into account. Figure \ref{fig:mvvs} considers several different scenarios for different values of the heavy scalar mass as well as mixing angles, where the pure SM-like contribution is subtracted. We see that for both scenarios discussed here the true invariant mass distribution, shown in blue, clearly deviates from the signal only distribution in black, in particular away from the resonance region where $m_{VV}\,=\,m_H$. In case the latter is assued, this would lead to clearly wrong distributions in such regions. For example, any neural network trained on such samples would then mispresent the correct distribution.

\begin{figure} 
\begin{center}
\includegraphics[width=0.9\textwidth]{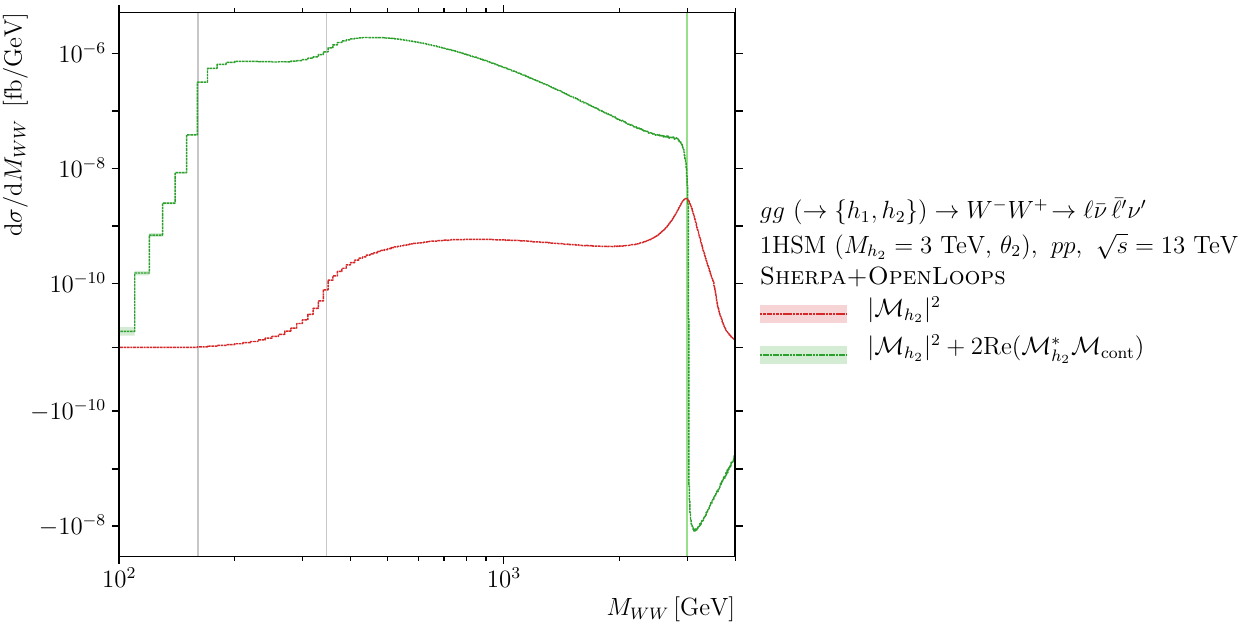}
\end{center}
\caption{\label{fig:invsherp} Invariant mass distribution in the $W^+W^-$ final state for a heavy resonance of 3 \TeV: signal only (red) as well as signal including interference with the continuum background (green). Figure is taken from \cite{Kauer:2019qei}. $h1/2$ denotes the lighter/ heavier scalar, and the mixing angle is here denoted by $\theta_2\,\approx\,0.13$. }
\end{figure}

In \cite{Kauer:2019qei}, the authors consider the same process in an OpenLoops+Sherpa framework \cite{Buccioni:2019sur,Sherpa:2024mfk}. In addition to invariant mass distributions, they also investigate the dependence of other variables that typically enter in cut determination for such searches, as e.g. the transverse mass, dilepton invariant masses or angular observables; we refer the reader to the original reference for more details. We here just display the invariant mass distribution from signal alone as well as signal and interference with the continuum contribution for a heavy resonance at 3 \TeV with an angle $\theta_2\,\sim\,0.13$ in figure \ref{fig:invsherp}\footnote{In \cite{Kauer:2019qei} different choices for the mixing angle in the singlet extension are denoted by $\theta_{1,2}$ which should be equated to $\al$ in the notation of this work. See also table \ref{tab:wwcontrs}.}.

In addition, the authors also quantify the magnitude of the different contributions in detailed numerical studies. An example is given in table \ref{tab:wwcontrs} which is taken from that reference and lists these contributions in great detail.

\begin{table}[tbp]
\vspace{0.cm}
\centering
\renewcommand{\arraystretch}{1.2}
\begin{tabular}{|c|c|c|c|c|}
\cline{2-5}
\multicolumn{1}{c|}{} & \multicolumn{4}{|c|}{$gg\ (\to \{h_1,h_2\}) \to W^-W^+ \!\to \ell\bar{\nu}\,\bar{\ell}^\prime \nu^\prime$} \\ 
\multicolumn{1}{c|}{} &   \multicolumn{4}{|c|}{$\sigma$ [fb], $pp$, $\sqrt{s}=13$ TeV} \\ 
\multicolumn{1}{c|}{} &   \multicolumn{4}{|c|}{1HSM} \\ 
\cline{2-5}
\multicolumn{1}{c|}{} & $M_{h_2}$ & \multicolumn{3}{|c|}{$|\mathcal{M}|^2$} \\
\cline{3-5}
\multicolumn{1}{c|}{} &  [GeV]  & Sq($h_2$) & $h_2$+I($h_1$) & $h_2$+I(C+$h_1$) \\
\hline
\multirow{8}{*}{$\theta_1$}
 & 700 & $0.07810(2)$ & $0.04113(4)$ & $0.09591(7)$ \\
 & ratio & $1$ & $0.5266(6)$ & $1.2280(9)$ \\ \cline{2-5}
 & 1000 & $0.010824(2)$ & $-0.01621(2)$ & $0.01780(3)$ \\
 & ratio & $1$ & $-1.498(2)$ & $1.644(2)$ \\ \cline{2-5}
 & 1500 & $0.00027818(5)$ & $-0.005749(2)$ & $0.001214(3)$ \\
 & ratio & $1$ & $-20.668(8)$ & $4.36(1)$ \\ \cline{2-5}
 & 3000 & $5.3026(9)\times 10^{-7}$ & $-0.00032008(7)$ & $3.46(2)\times 10^{-5}$ \\
 & ratio & $1$ & $-603.6(2)$ & $65.2(2)$ \\
\hline
\multirow{8}{*}{$\theta_2$}
 & 700 & $0.27776(5)$ & $0.1737(2)$ & $0.3502(2)$ \\
 & ratio & $1$ & $0.6252(4)$ & $1.2606(7)$ \\ \cline{2-5}
 & 1000 & $0.035182(6)$ & $-0.03845(3)$ & $0.06833(5)$ \\
 & ratio & $1$ & $-1.0928(9)$ & $1.942(2)$ \\ \cline{2-5}
 & 1500 & $0.0008885(2)$ & $-0.016227(5)$ & $0.005293(7)$ \\
 & ratio & $1$ & $-18.262(6)$ & $5.957(8)$ \\ \cline{2-5}
 & 3000 & $2.3605(4)\times 10^{-6}$ & $-0.0010870(3)$ & $0.0001561(4)$ \\
 & ratio & $1$ & $-460.5(2)$ & $66.1(2)$ \\
\hline
\end{tabular}\\[0cm] 
\caption{\label{tab:wwcontrs}
Cross sections for $gg\ (\to \{h_1,h_2\}) \to W^-W^+ \!\to \ell\bar{\nu}\,\bar{\ell}^\prime \nu^\prime$ in $pp$ collisions at $\sqrt{s}=13$~TeV in the 1HSM with focus on heavy Higgs ($h_2$) production. $Sq (h_2)$ denotes the contribution from the signal process, while the last two colums give the magnitude of the interference with the SM-like scalar and the same including continuum, respectively. The angles are mass dependent and vary between 0.07 for 3 \TeV and 0.39 for 700 \GeV; see original work for details. The ratio $\sigma/\sigma(\mathrm{Sq}(h_2))$ is also given. Table including caption taken from \cite{Kauer:2019qei}.}
\end{table}

\begin{figure}
\begin{center}
\includegraphics[width=0.75\textwidth]{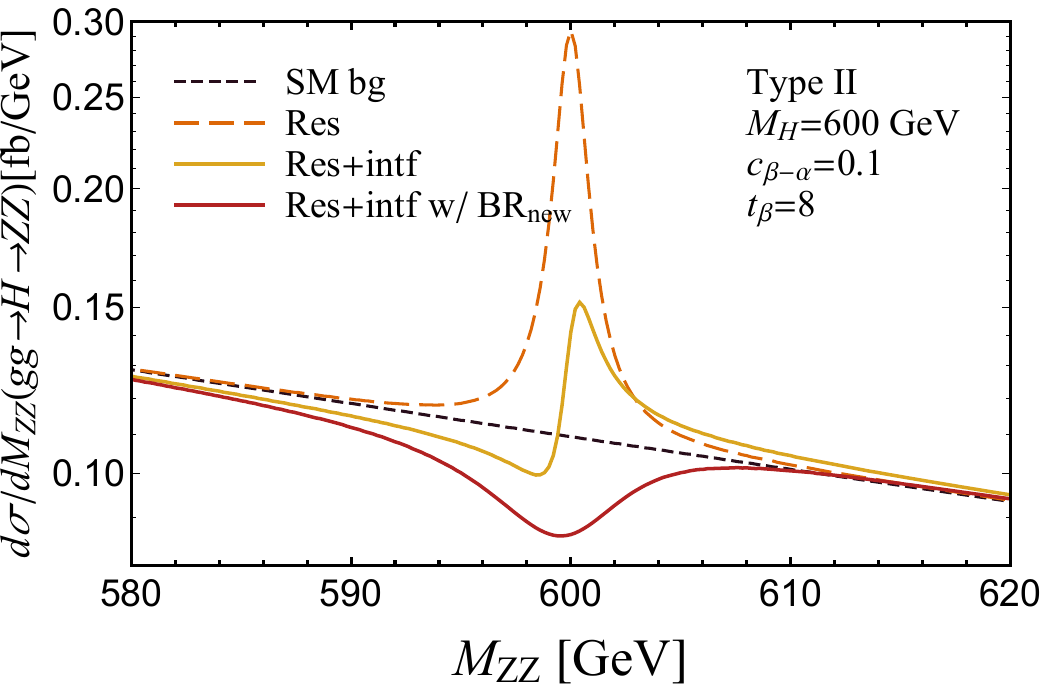}

\caption{\label{fig:zz2hdm} Invariant mass distributions for the diboson system in a 2HDM with parameters as specified in the caption, for various width assumptions. It is clear that the resonance only distribution does not give the full picture. We also see a clear modification when the width is varied. Taken from \cite{Jung:2015sna}.}
\end{center}
\end{figure}

Such final states have also been studied in other new physics scenarios. In figure \ref{fig:zz2hdm}, we show results that were e.g. derived in \cite{Jung:2015sna} in the context of a 2HDM. Again we display the invariant mass distribution for the diboson final state for resonance only, resonance including interference with the continuum, as well as the resonance assuming that there are new final states contributing to the decay width of $H$ with $\text{BR}_\text{new}\,=\,0.8$. As before, it becomes obvious that taking the resonance distribution alone clearly does not model the diboson invariant mass correctly.

Finally, we want to mention some work where the decay into di-gluon was investigated \cite{Martin:2016bgw,Bhattiprolu:2020zoq}. Again, the authors observe significant distortions in the di-jet invariant mass distributions with respect to the naively expected Breit-Wigner shape.

\subsection{Di-Higgs final states}
We now turn to di-Higgs production processes. The process  $p\,p\,\rightarrow\,h\,h$ is currently of high interest, as it allows in principle for the determination of the triple scalar coupling $\lam_{hhh}$ which in turn can give insight into the evolution of the universe e.g. in the form of electroweak phase transitions. 

While the SM non-resonant final states are still elusive, resonance-enhanced processes can render much higher rates than pure SM predictions, which are on the order of 30-40 \fb for LHC Run 3 and HL-LHC center-of-mass energies \cite{Grazzini:2018bsd}\footnote{See also \cite{Baglio:2020wgt} for a related uncertainty discussion.}, and therefore the LHC experiments can also set important limits for such processes. However, again it is of high importance to model the new physics contributions in a correct way such that the experimental results can correctly be translated into bounds on an underlying UV-complete model.

\begin{figure}
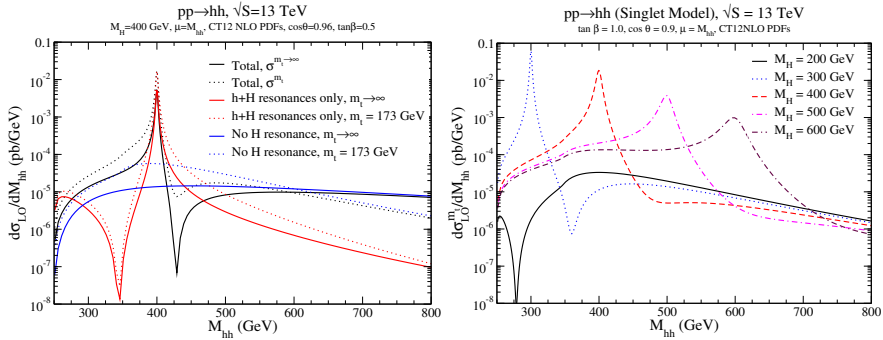
  
\begin{center}
\begin{minipage}{0.45\textwidth}
\includegraphics[width=\textwidth]{res_400_tb5}
\end{minipage}
\begin{minipage}{0.45\textwidth}
\includegraphics[width=\textwidth]{dsig_singlet_masses_tb1_new}
\end{minipage}
\end{center}
\caption{\label{fig:mhhdists} Example for singlet extension. Figure highlightening the effects of including interference contributions. {\sl Left:} Different contributions for a fixed new physics scenario, both in the full mass dependence and heavy top limit. {\sl Right:} Total invariant mass distribution including interference effects for various new physics scenarios. Figures are taken from \cite{Dawson:2015haa}. Here $h/H$ denotes the light/ heavy resonance. }
\end{figure}

The effects of including interference terms in invariant mass distributions for di-Higgs final states induced by heavy resonances have been known for quite some time, see e.g. \cite{Dawson:2015haa,Carena:2018vpt,DiMicco:2019ngk}, with some interesting initial work presented e.g. in \cite{Dawson:2015haa}. In figure \ref{fig:mhhdists}, we display the invariant diboson mass distribution for various scenarios. While the left part of the plot displays different contributions to the total final state, the right part of the figure displays the total contributions taking all interference effects into account for different new physics scenarios. We see that for all cases the invariant mass distributions do not correspond to a pure Breit-Wigner description, but differ due to interference effects.

\begin{figure}
  \begin{center}
 \includegraphics[width=0.35\textwidth]{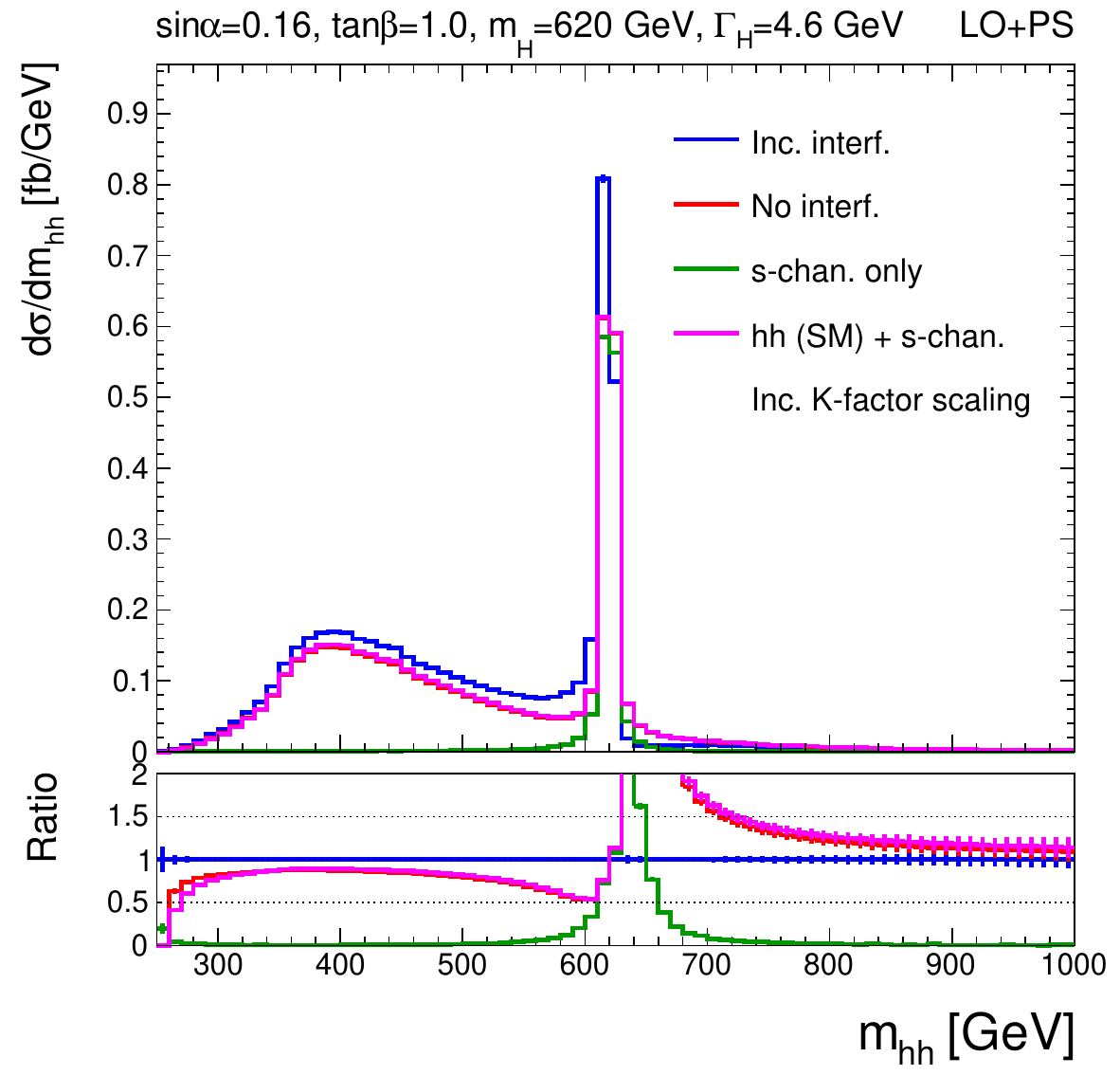}
 \includegraphics[width=0.35\textwidth]{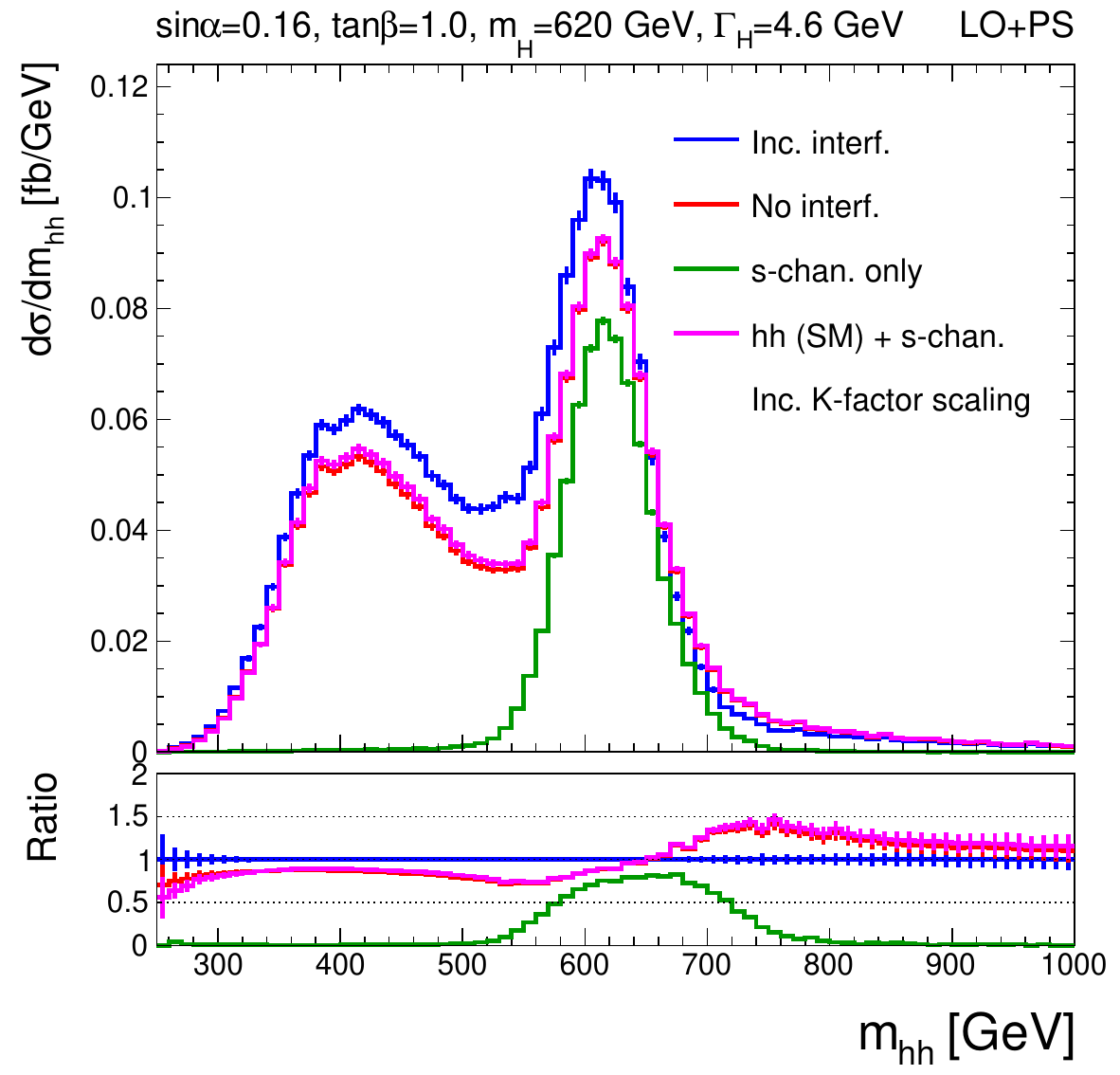}\\
 \includegraphics[width=0.35\textwidth]{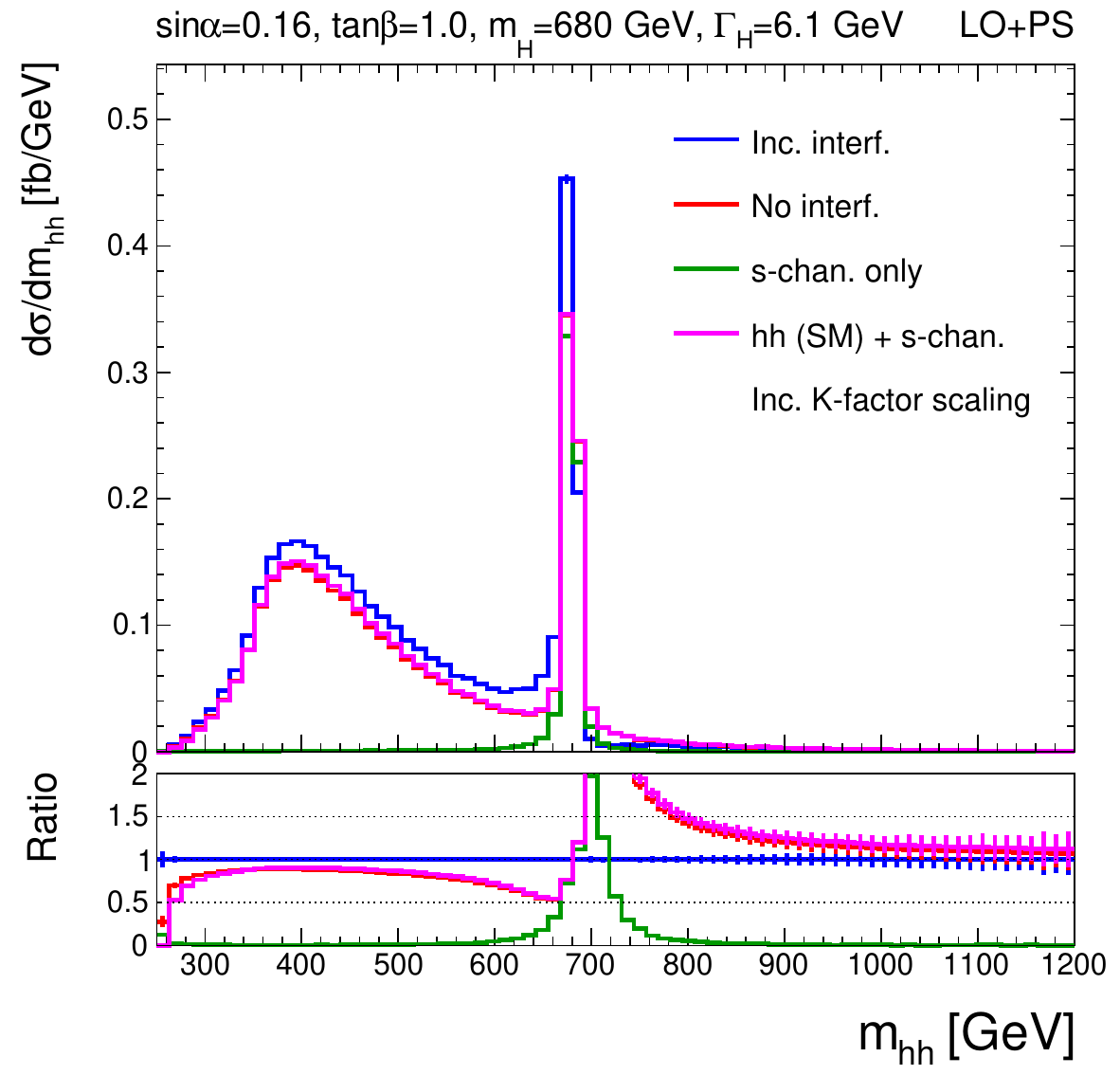}
 \includegraphics[width=0.35\textwidth]{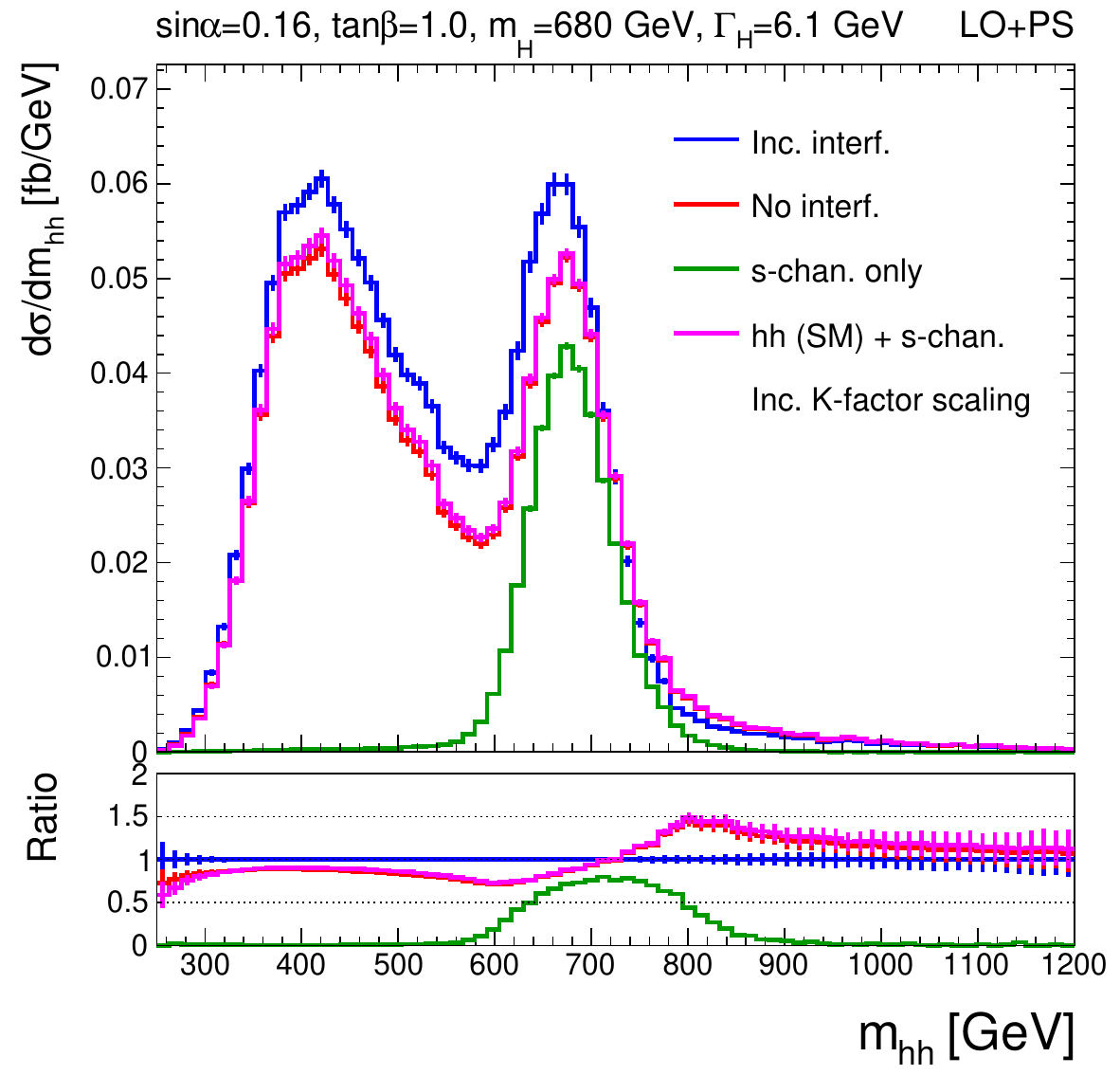}
 \caption{\label{fig:ourbps} Example for singlet extension. Invariant mass distributions for different benchmark scenarios for resonance enhanced di-Higgs production. {\sl Left:} Prior to and {\sl right} after taking smearing effects into account, resulting from finite detector resolution. Displayed are the pure resonances {\sl (green)}, SM background and resonance contribution with {\sl (red)} and without {\sl (pink)} correct coupling rescalings of the latter, as well as full contribution including all interference terms {\sl (blue)}. Figures are taken from \cite{Feuerstake:2024uxs} and include K-factors that normalize to the total production cross sections at next-to-leading order (NLO) (see original reference for details). $h/H$ denote the light/ heavy resonance. }
  \end{center}
\end{figure}

As mentioned before, it is also quite important to study these effects for other kinematic variables that might be used e.g. to design cuts for experimental searches, and also to include realistic smearing from detector effects. Both points are considered in \cite{Feuerstake:2024uxs}, where in addition a number of benchmark points have been identified that are used to display different effects that can occur in these benchmark points as e.g. maximal interference effects or maximal triple scalar couplings. Two of these benchmark points are displayed in figure \ref{fig:ourbps}, prior and after smearing effects from finite detector resolution are taken into account.

In particular after smearing, the invariant mass distributions stemming from the SM-like contributions can mimick a second mass peak. In addition, the interference terms for these benchmark points can lead to rate enhancements in the peak as well as offshell regions. More recent related theory studies can be found e.g. in \cite{Arco:2025nii,Moretti:2025dfz,Braathen:2025svl,Hammad:2025aka}.

\begin{figure}
\begin{center}
  \includegraphics[width=0.9\textwidth]{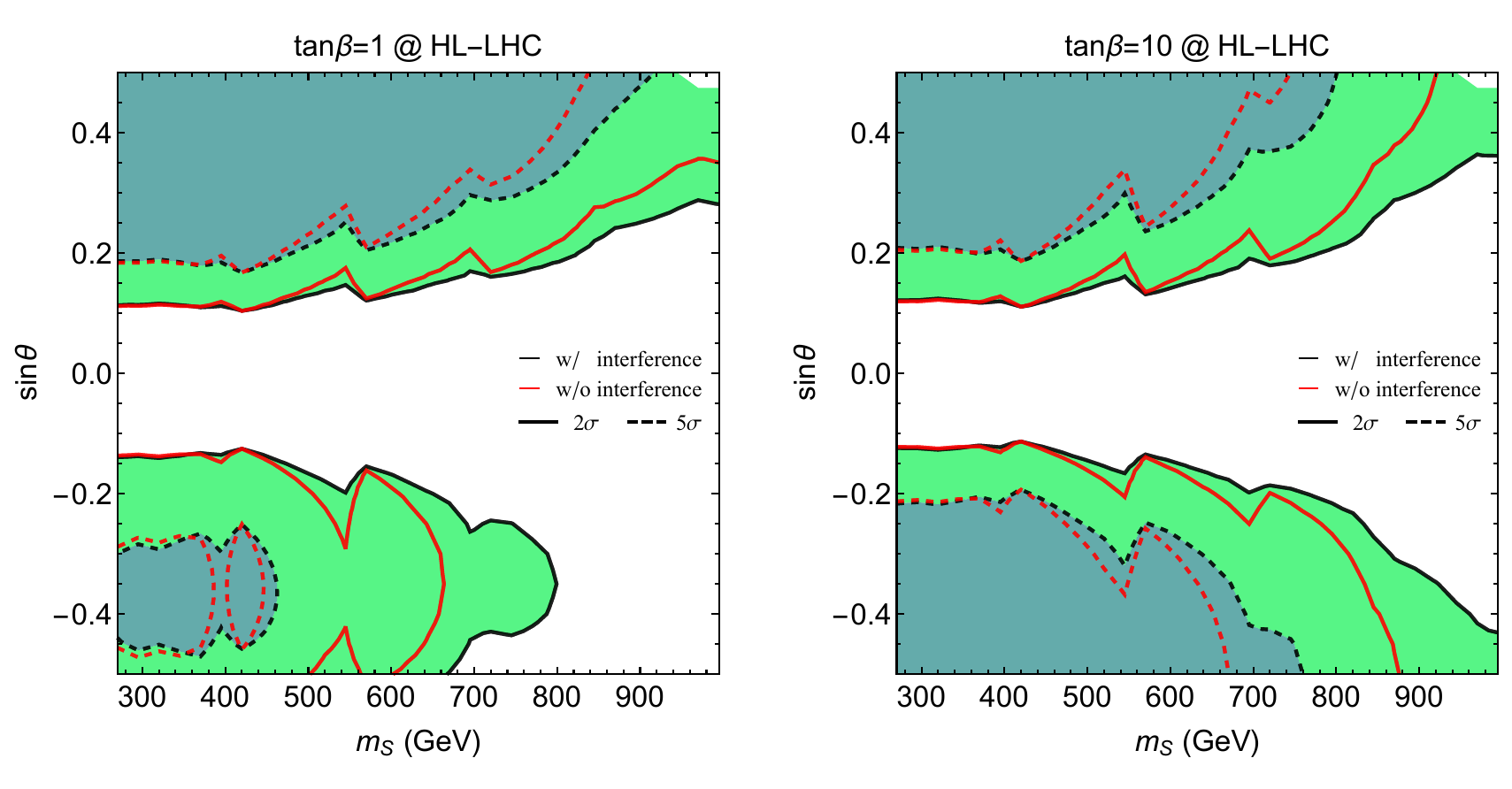}
  \caption{\label{fig:hlcarena} Projections for both exclusion and discovery projections at the HL-LHC for a singlet extension for di-Higgs final states where the singlet acquires a vev, for different values of the additional vev (here parametrized via $\tan\be$). In both scenarios, there can be significant differences between ignoring and including the interferences. Figure is taken from \cite{Carena:2018vpt}. $S$ here denotes the heavy scalar resonance.}
\end{center}
\end{figure}

In addition to the differences in differential distributions, one can also investigate the effect of including or not including interference terms in the search strategies. This has been done again for a simple singlet extension in \cite{Carena:2018vpt}. We display the results of such a comparison in figure \ref{fig:hlcarena}, where the authors have projected the discovery reach with and without taking interference contributions into account. They study two different scenarios for the ratio of the two vacuum expectation values and explore the respective reach at the HL-LHC. In both cases, there can be significant differences depending on the values of the additional new physics parameters.

\begin{figure}
  \begin{center}
\includegraphics[width=0.6\textwidth]{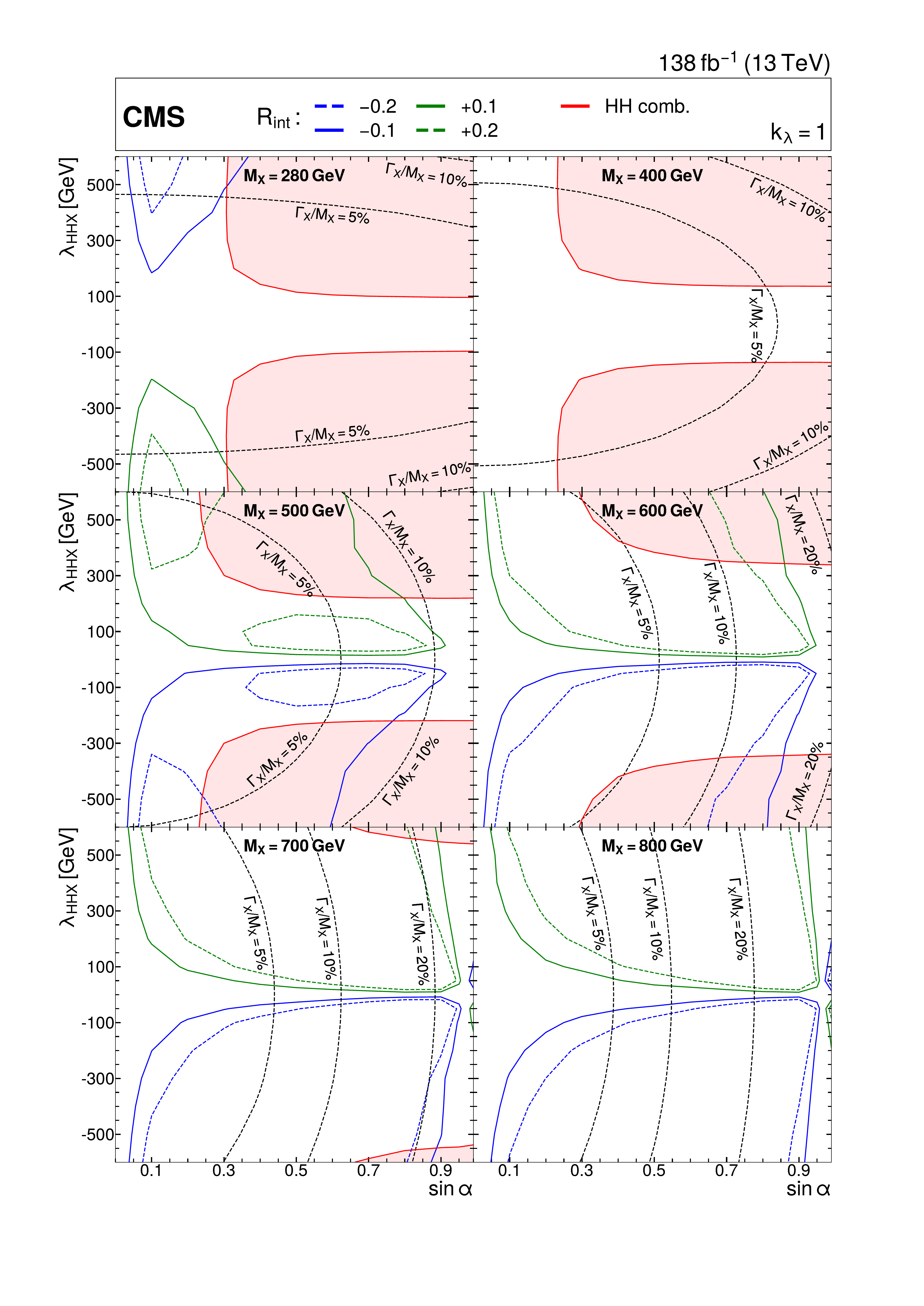}
\caption{\label{fig:if_exp} Figure from the CMS Run 2 combination \cite{CMS:2024phk}, for the scenario where the SM scalar sector is enhanced by an additional heavy resonance, denoted by $X$, and rescaled couplings similar to the singlet scenario. Shown are scenarios for different heavy resonance masses, derived bounds on the $X\,H\,H$ couplings $\lam_{HHX}$ as a function of the mixing angle, the ratio of width over mass that can serve as a quantifier for the NWA, as well as values of $R$ defined in eqn. (\ref{eq:rif}). Here, $X$ signifies the heavy resonance decaying into di-Higgs final states. Regions that are currently excluded using only the signal, i.e. with larger values of $|R|$, should be reinvestigated using the full simulation. }
  \end{center}
\end{figure}

Along these lines, a first attempt has been made by the CMS collaboration to at least quantify where interference effects could become important \cite{CMS:2024phk}. This work discusses various possible discovery channels including Higgs bosons in the final states, the status after Run 2 as well as a possible projection for HL-LHC for these searches. For the di-Higgs final states stemming from heavy resonances, the collaboration defined a ratio
\begin{\eqn}\label{eq:rif}
  R\,=\,\frac{\sigma_\text{full}-\lb\sigma_{H}+\sigma_{no H} \rb}{\sigma_{H}+\sigma_{no H}}
  \end{\eqn}
in order to quantify the interference effects for $H\,H$ final states\footnote{It would be more advisable to change the denominator to $\sigma_\text{full}$ for a complete understanding of the overall magnitude of such effects.}. We display the corresponding contours in figure \ref{fig:if_exp}.

Finally, di-Higgs invariant mass distributions can also be modified by including higher-order corrections to the scalar couplings. However, this is beyond the scope of the current review. We refer the interested reader to e.g. \cite{Heinemeyer:2024hxa} for a corresponding discussion.

\subsection{Triple Higgs final states}

Another interesting quantity in the scalar sector that determines the form of the potential and thereby evolution of the phase transition is the quartic coupling. This coupling is e.g. accessible in the measurement of three Higgs bosons in the final state, see e.g. \cite{Abouabid:2024gms} for a recent overview.

Recently, the first experimental results \cite{ATLAS:2024xcs,ATLAS:2025mim,CMS-PAS-HIG-24-012} for this process have become available. Here,  \cite{ATLAS:2024xcs,ATLAS:2025mim} study resonance-enhanced triple Higgs production via the chain

\begin{figure}
\begin{center}
\includegraphics[width=0.4 \textwidth]{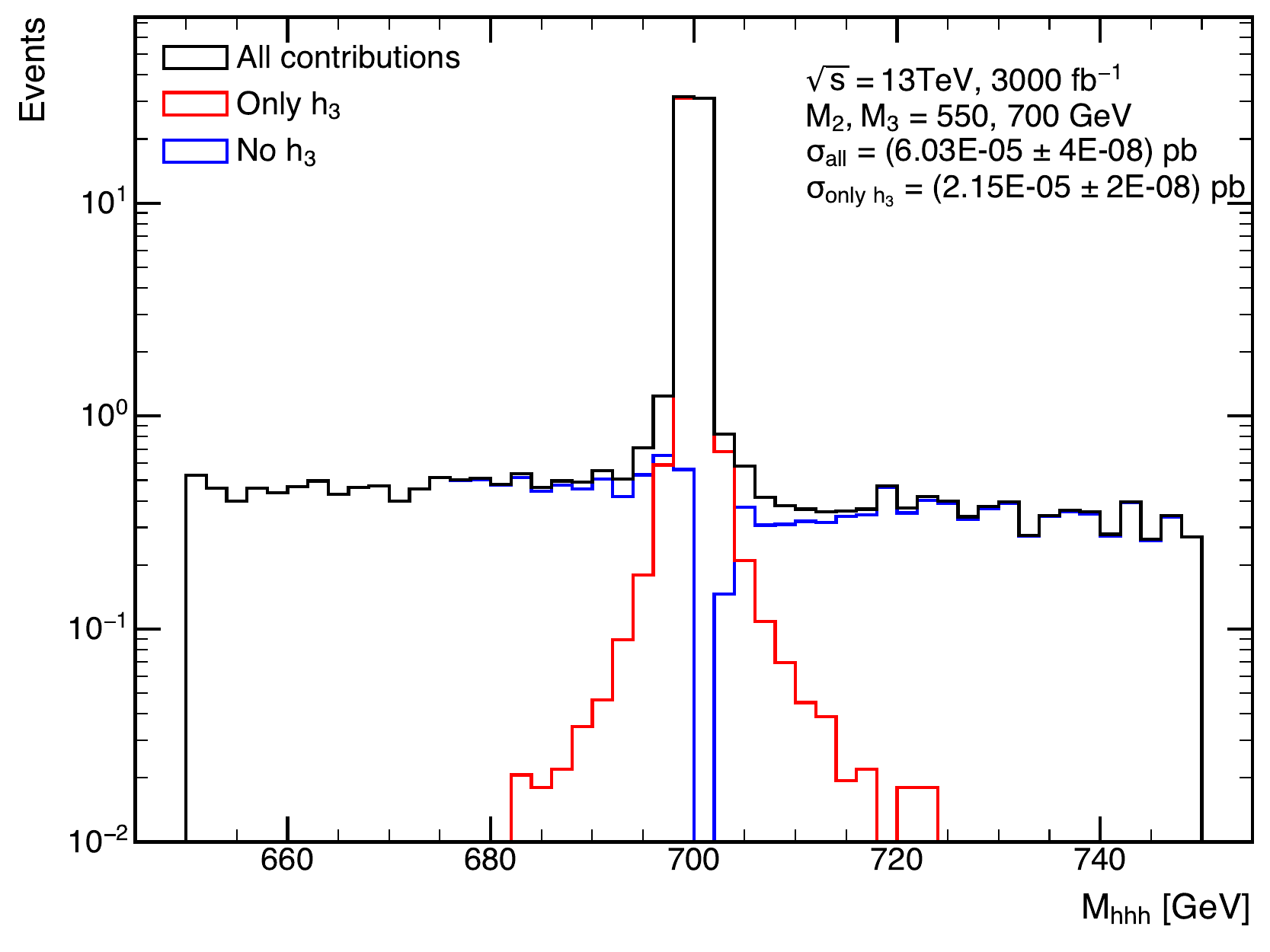}
\includegraphics[width=0.4 \textwidth]{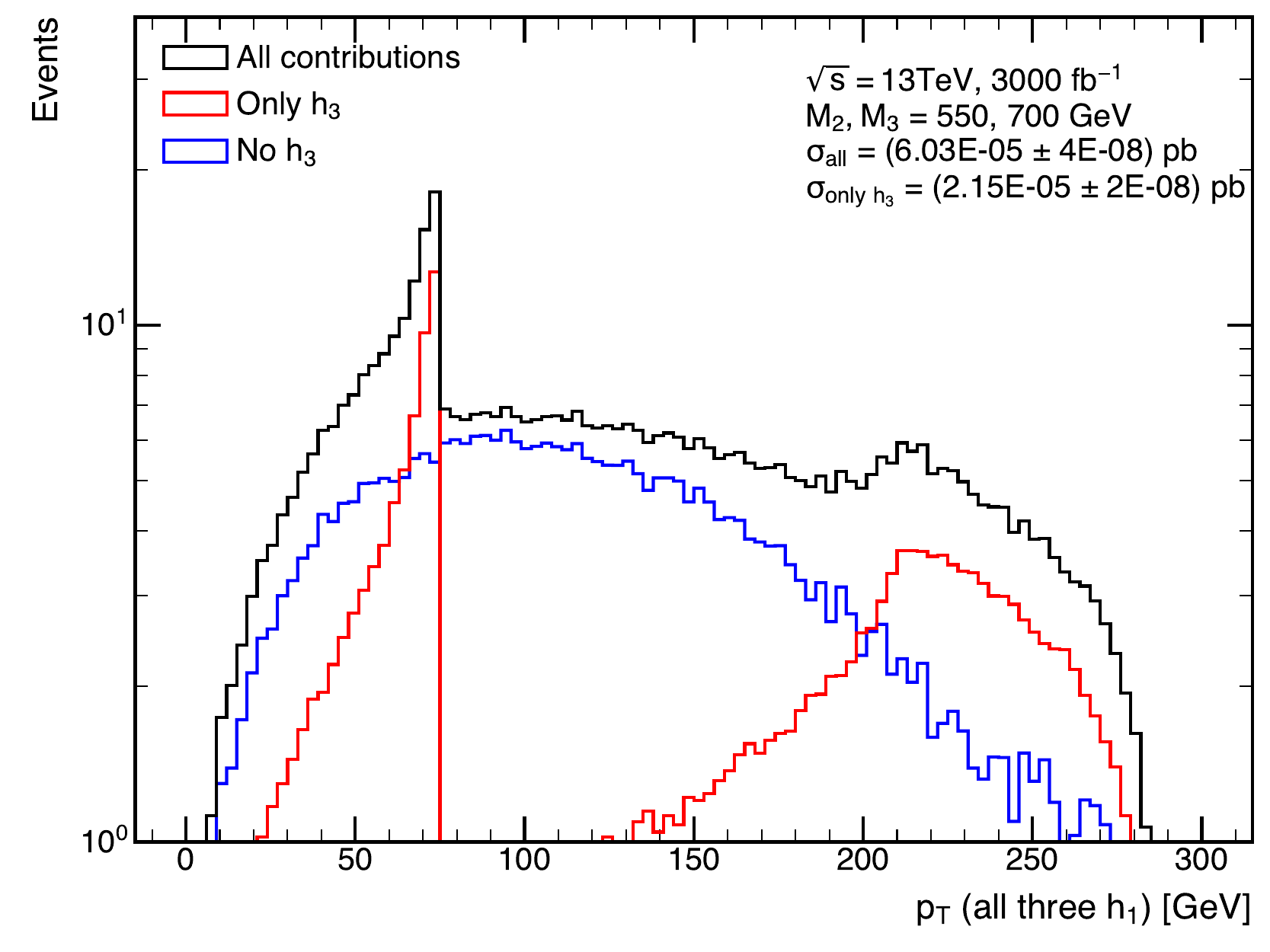}
\caption{\label{fig:hhh} Invariant triple Higgs mass distribution {\sl (left)} as well as $p_\perp$ distribution for all scalars {\sl (right)} of the process given in Eqn. (\ref{eqn:hhh}), where we display the signal only contribution {\sl (red)}, the total process without signal contribution {\sl (blue)}, and all contributions where no intermediate state has been specified {\sl (black)}. While the invariant mass seems to be well-described by the signal only contribution in the resonant region, it is clear that a full description calls for including all contributions. Figure is taken from \cite{Abouabid:2024gms}.}

\end{center}
\end{figure}

\begin{\eqn}\label{eqn:hhh}
  p\,p\,\rightarrow\,h_3\,\rightarrow\,h_2\,h_1\,\rightarrow\,h_1\,h_1\,h_1,
\end{\eqn}
where $h_1$ now takes the role of the 125 \GeV~ resonance and $h_{2/3}$ denote additional scalar states at higher masses. Note for scenarios like this there need to be at least 3 neutral final states, as e.g. in the model discussed in \cite{Robens:2019kga,Robens:2022nnw}. Figure \ref{fig:hhh} shows the triple Higgs invariant mass as well as the transverse momenta of all Higgs bosons, with either only the signal contribution, only background, or all contributions including interferences. While the invariant mass seems to be relatively well modelled by the signal only in the resonance region, it is clear that the full signal and background, including interference terms, needs to be taken into account to correctly model all distributions. For $p_T$, on the other hand, it is clear that including the signal only would clearly mismodel the actual distributions, in particular in an intermediate range. It has to be reemphasized that this is particular to the benchmark studied here and might differ for other scenarios. These results have been first presented in \cite{Abouabid:2024gms}\footnote{The author thanks R. Zhang for fruitful collaboration on this topic.}.

\subsection{Difermion final states: $t\bar{t}$}

\begin{figure}
\begin{center}
\includegraphics[width=0.5\textwidth]{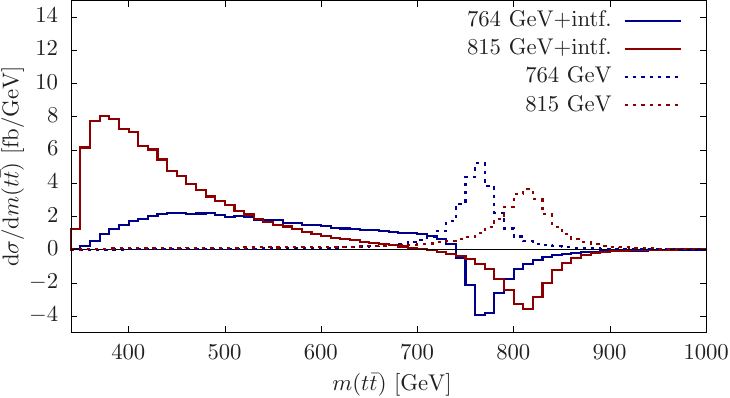}
\includegraphics[width=0.4\textwidth]{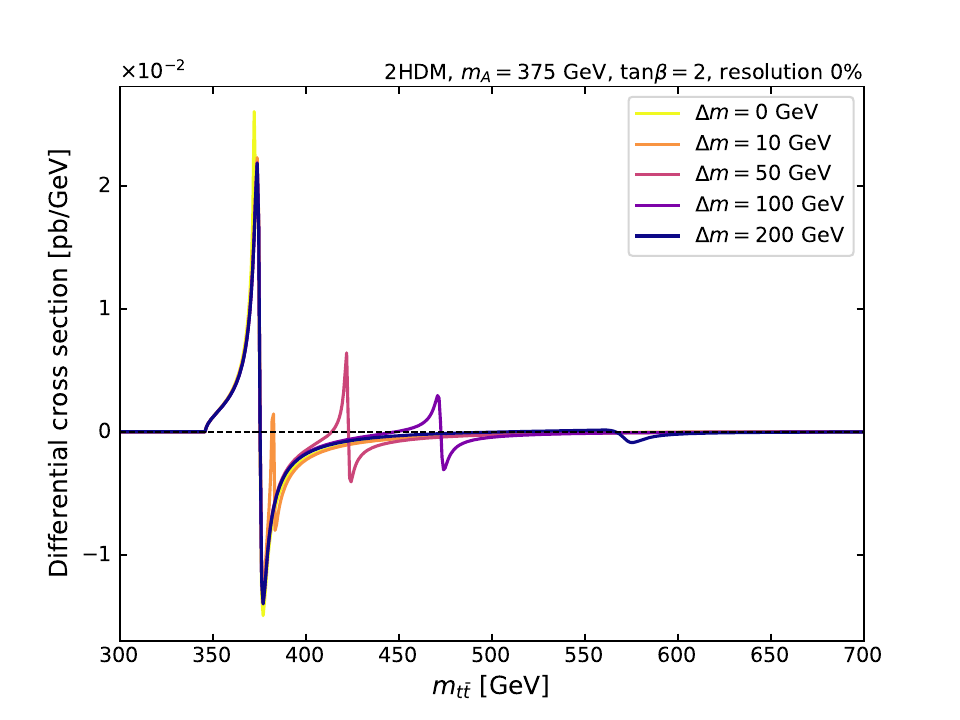}
\caption{\label{fig:ttfin} Interference effects in the di-top invariant mass distributions in 2HDMs, both background-subtracted. {\sl Left:} signal-only and signal + interference with the background in a complex 2HDM with two different resonances as indicated in the figure, taken from \cite{Basler:2019nas}. {\sl Right:} Invariant di-top mass distributions for a 2HDM with different mass-degeneracies between the two different additional scalars ranging from $0\GeV$ {\sl (yellow)} to $200\,\GeV$ {\sl (dark blue)}. Away from the resonance region, different mass differences lead to different distortions in the invariant mass spectra. This figure assumes an ideal detector-resolution of $0\%$; results including smearing can be found in the original work. Figure taken from \cite{Djouadi:2019cbm}.}
\end{center}
\end{figure}

Also for processes where the additional new physics state can decay primarily into fermionic final states, interference effects can become important. A typical scenario where such processes can occur are two Higgs doublet models discussed above.
In such models the new scalar states are often relatively mass-degenerate, which is due to a combination of theoretical constraints on the couplings as well as bounds from electroweak precision observables. In figure \ref{fig:ttfin}, we show results from \cite{Basler:2019nas,Djouadi:2019cbm}, again with a focus on the di-fermion invariant mass distributions and subtracting the pure SM-like background. As before, it becomes obvious that instead of a pure Breit-Wigner distribution around the resonance peak, taking into account interference effects significantly distorts the respective distributions. We refer the reader to the original publications for model specifications.

\begin{figure}
    \centering
    \includegraphics[width=0.5\linewidth]{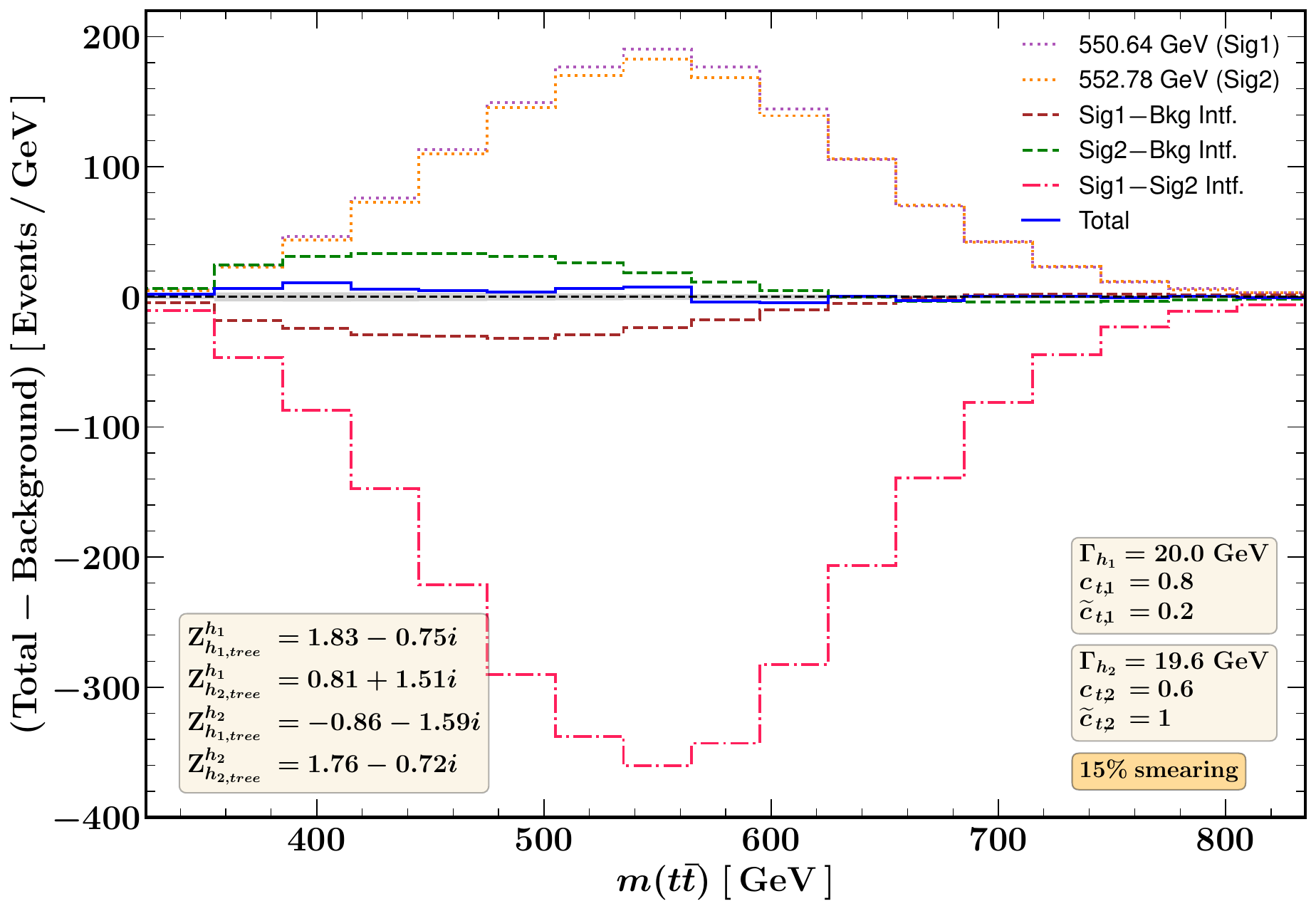}
    \caption{\label{fig:ttgeorg}  Scenario involving large cancellations: contributions to the background-subtracted invariant mass distribution at the hadronic level for $t\,\bar{t}$ production. The tree-level masses are $550$ GeV and $571$ GeV. All the other parameters are indicated in the plot. Figure taken from \cite{Bahl:2025you}.}
\end{figure}

Finally, we showcase an example in $t\,\bar{t}$ production where two resonances interfere in such a way that the respective contributions to the di-top invariant mass cancel, as discussed e.g. by the authors of \cite{Bahl:2025you}. They discuss $t\,\bar{t}$ production in the scenario of a complex 2HDM with two nearly mass-degenerate additional scalar states that interfere in a negative way. The resulting invariant mass distribution is shown in figure \ref{fig:ttgeorg}.

After all contributions, including interference effects, are taken into account in fact the subtracted invariant mass distribution turns out as a flat plateau, despite the presence of two resonances that contribute to $t\,\bar{t}$ production in the s-channel. Although this can be considered an extreme scenario, it demonstrates the fact that interferences are important in realistic models and need to be taken into account for a correct modelling of UV complete physics scenarios. In the same work, the authors also showcase a model where, although nonzero, the invariant mass distribution lies within the experimental errorband for certain regions in phase space.

\section{Conclusion}
In this short review, I have emphasized the importance of interference effects in searches for new physics scenarios that render an s-channel resonance decaying into SM-like final states. For most of the examples shown here, the interference with either the SM contribution or additional scalar states leads to a clear modification of the respective invariant mass distributions away from the expected Breit-Wigner shape around the peak. This effect can also be observed in other quantities as e.g. transverse momenta of the decay products. Therefore, realistic searches for such resonances need to include these effects, in particular when neural networks or other advanced tools are used for cut optimization. Leading-order tools to model these scenarios are readily available. The correct treatment of higher-order corrections in the production mechanism is still an open topic and can currently only be approximated if no dedicated NLO treatment is available, but proposals to treat these are readily available in the literature. We therefore encourage the experimental collaborations to include these effects in their studies. A common prescription that is already applied by the collaborations in such scenarios is e.g. event reweighting. Alternatively appropriate simulations should be used.

\section*{Acknowledgements}
TR acknowledges financial support from the Croatian Science Foundation (HRZZ) project " Beyond the Standard Model discovery and Standard Model precision at LHC Run III", IP-2022-10-2520. I also thank the theory group of the Max-Planck-Institute for Nuclear Physics in Heidelberg for the invitation to their seminar, which triggered the invite to submit this review. I also want to thank F. Feuerstake, E. Fuchs and D. Winterbottom for fruitful collaboration on a related work \cite{Feuerstake:2024uxs}.

\bibliography{lit.bib}

@article{Martin:2016bgw,
    author = "Martin, Stephen P.",
    title = "{Signal-background interference for a singlet spin-0 digluon resonance at the LHC}",
    eprint = "1606.03026",
    archivePrefix = "arXiv",
    primaryClass = "hep-ph",
    doi = "10.1103/PhysRevD.94.035003",
    journal = "Phys. Rev. D",
    volume = "94",
    number = "3",
    pages = "035003",
    year = "2016"
}

@article{Bhattiprolu:2020zoq,
    author = "Bhattiprolu, Prudhvi N. and Martin, Stephen P.",
    title = "{Signal-background interference for digluon resonances at the Large Hadron Collider}",
    eprint = "2004.06181",
    archivePrefix = "arXiv",
    primaryClass = "hep-ph",
    doi = "10.1103/PhysRevD.102.015016",
    journal = "Phys. Rev. D",
    volume = "102",
    number = "1",
    pages = "015016",
    year = "2020"
}

@techreport{CMS-PAS-HIG-24-012,
      collaboration = "CMS",
      title         = "{Search for nonresonant triple Higgs boson production in
                       the six b-quark final state in proton-proton collisions at
                       13 TeV}",
      institution   = "CERN",
      reportNumber  = "CMS-PAS-HIG-24-012",
      address       = "Geneva",
      year          = "2025",
      note          = "CMS-PAS-HIG-24-012",
      url           = "https://cds.cern.ch/record/2945361",
}

@article{Bahl:2025you,
    author = "Bahl, Henning and Kumar, Romal and Weiglein, Georg",
    title = "{Impact of interference effects on Higgs-boson searches in the di-top final state at the LHC}",
    eprint = "2503.02705",
    archivePrefix = "arXiv",
    primaryClass = "hep-ph",
    reportNumber = "DESY-25-030",
    doi = "10.1007/JHEP05(2025)098",
    journal = "JHEP",
    volume = "05",
    pages = "098",
    year = "2025"
}

@article{Moretti:2025dfz,
    author = {Moretti, Stefano and Panizzi, Luca and Sj{\"o}lin, J{\"o}rgen and Waltari, Harri},
    title = "{Deconstructing resonant Higgs boson pair production at the LHC: Effects of colored and neutral scalars in the NMSSM test case}",
    eprint = "2506.09006",
    archivePrefix = "arXiv",
    primaryClass = "hep-ph",
    doi = "10.1103/p2p4-67g9",
    journal = "Phys. Rev. D",
    volume = "112",
    number = "5",
    pages = "055005",
    year = "2025"
}

@article{Arco:2025nii,
    author = {Arco, F. and Heinemeyer, S. and M{\"u}hlleitner, M. and Arnay, A. Parra and Gonz{\'a}lez, N. Rivero and Schaeidt, A. Verduras},
    title = "{Sensitivity to triple Higgs couplings via di-Higgs production in the RxSM at the (HL-)LHC and future e$^{+}$e$^{-}$ colliders}",
    eprint = "2502.03878",
    archivePrefix = "arXiv",
    primaryClass = "hep-ph",
    reportNumber = "DESY-24-213, IFT--UAM/CSIC-24-186, KA-TP-01-2025",
    doi = "10.1007/JHEP06(2025)211",
    journal = "JHEP",
    volume = "06",
    pages = "211",
    year = "2025"
}

@article{Braathen:2025svl,
    author = "Braathen, Johannes and Heinemeyer, Sven and Boatella, Carlos Pulido and Verduras Schaeidt, Alain",
    title = "{Complementarity of gravitational wave analyses and di-Higgs production in the exploration of the Electroweak Phase Transition dynamics in the RxSM}",
    eprint = "2510.12569",
    archivePrefix = "arXiv",
    primaryClass = "hep-ph",
    reportNumber = "DESY-25-136, IFT--UAM/CSIC-25-110",
    month = "10",
    year = "2025"
}

@article{Hammad:2025aka,
    author = "Hammad, A. and Moretti, S. and Przybyl, A. P. and Waltari, H.",
    title = "{Interference Effects in Resonant Standard Model di-Higgs Production and Decay into $4b$ Final States: the Role of Machine Learning Analysis}",
    eprint = "2512.12318",
    archivePrefix = "arXiv",
    primaryClass = "hep-ph",
    month = "12",
    year = "2025"
}

@article{CMS:2024phk,
    author = "Hayrapetyan, Aram and others",
    collaboration = "CMS",
    title = "{Searches for Higgs boson production through decays of heavy resonances}",
    eprint = "2403.16926",
    archivePrefix = "arXiv",
    primaryClass = "hep-ex",
    reportNumber = "CMS-B2G-23-002, CERN-EP-2024-062",
    doi = "10.1016/j.physrep.2024.09.004",
    journal = "Phys. Rept.",
    volume = "1115",
    pages = "368--447",
    year = "2025"
}

@article{Jung:2015sna,
    author = "Jung, Sunghoon and Yoon, Yeo Woong and Song, Jeonghyeon",
    title = "{Interference effect on a heavy Higgs resonance signal in the $\gamma \gamma$ and $ZZ$ channels}",
    eprint = "1510.03450",
    archivePrefix = "arXiv",
    primaryClass = "hep-ph",
    reportNumber = "KIAS-P15055",
    doi = "10.1103/PhysRevD.93.055035",
    journal = "Phys. Rev. D",
    volume = "93",
    number = "5",
    pages = "055035",
    year = "2016"
}

@article{DiMicco:2019ngk,
    author = "Alison, J. and others",
    editor = "Di Micco, Biagio and Gouzevitch, Maxime and Mazzitelli, Javier and Vernieri, Caterina",
    title = "{Higgs boson potential at colliders: Status and perspectives}",
    eprint = "1910.00012",
    archivePrefix = "arXiv",
    primaryClass = "hep-ph",
    reportNumber = "FERMILAB-CONF-19-468-E-T, LHCXSWG-2019-005",
    doi = "10.1016/j.revip.2020.100045",
    journal = "Rev. Phys.",
    volume = "5",
    pages = "100045",
    year = "2020"
}

@article{ATLAS:2024xcs,
    author = "Aad, Georges and others",
    collaboration = "ATLAS",
    title = "{Search for triple Higgs boson production in the 6b final state using pp collisions at s=13{\,}{\,}TeV with the ATLAS detector}",
    eprint = "2411.02040",
    archivePrefix = "arXiv",
    primaryClass = "hep-ex",
    reportNumber = "CERN-EP-2024-285",
    doi = "10.1103/PhysRevD.111.032006",
    journal = "Phys. Rev. D",
    volume = "111",
    number = "3",
    pages = "032006",
    year = "2025"
}

@article{Robens:2019kga,
    author = "Robens, Tania and Stefaniak, Tim and Wittbrodt, Jonas",
    title = "{Two-real-scalar-singlet extension of the SM: LHC phenomenology and benchmark scenarios}",
    eprint = "1908.08554",
    archivePrefix = "arXiv",
    primaryClass = "hep-ph",
    reportNumber = "DESY-19-142, DESY 19-142",
    doi = "10.1140/epjc/s10052-020-7655-x",
    journal = "Eur. Phys. J. C",
    volume = "80",
    number = "2",
    pages = "151",
    year = "2020"
}

@article{Robens:2022nnw,
    author = "Robens, Tania",
    title = "{Two-Real-Singlet-Model Benchmark Planes}",
    eprint = "2209.10996",
    archivePrefix = "arXiv",
    primaryClass = "hep-ph",
    reportNumber = "RBI-ThPhys-2022-35",
    doi = "10.3390/sym15010027",
    journal = "Symmetry",
    volume = "15",
    pages = "27",
    year = "2023"
}

@article{Abouabid:2024gms,
    author = "Abouabid, Hamza and others",
    title = "{HHH whitepaper}",
    eprint = "2407.03015",
    archivePrefix = "arXiv",
    primaryClass = "hep-ph",
    doi = "10.1140/epjc/s10052-024-13376-3",
    journal = "Eur. Phys. J. C",
    volume = "84",
    pages = "1183",
    year = "2024"
}

@article{Djouadi:2019cbm,
    author = "Djouadi, Abdelhak and Ellis, John and Popov, Andrey and Quevillon, J{\'e}r{\'e}mie",
    title = "{Interference effects in $ t\overline{t} $ production at the LHC as a window on new physics}",
    eprint = "1901.03417",
    archivePrefix = "arXiv",
    primaryClass = "hep-ph",
    reportNumber = "CERN-TH-2019-001, KCL-PH-TH/2018-05, LAPTH/001/19",
    doi = "10.1007/JHEP03(2019)119",
    journal = "JHEP",
    volume = "03",
    pages = "119",
    year = "2019"
}

@article{Basler:2019nas,
    author = {Basler, Philipp and Dawson, Sally and Englert, Christoph and M{\"u}hlleitner, Margarete},
    title = "{Di-Higgs boson peaks and top valleys: Interference effects in Higgs sector extensions}",
    eprint = "1909.09987",
    archivePrefix = "arXiv",
    primaryClass = "hep-ph",
    doi = "10.1103/PhysRevD.101.015019",
    journal = "Phys. Rev. D",
    volume = "101",
    number = "1",
    pages = "015019",
    year = "2020"
}

@article{Heinemeyer:2024hxa,
    author = {Heinemeyer, S. and M{\"u}hlleitner, M. and Radchenko, K. and Weiglein, G.},
    title = "{Higgs pair production in the 2HDM: impact of loop corrections to the trilinear Higgs couplings and interference effects on experimental limits}",
    eprint = "2403.14776",
    archivePrefix = "arXiv",
    primaryClass = "hep-ph",
    doi = "10.1140/epjc/s10052-025-14124-x",
    journal = "Eur. Phys. J. C",
    volume = "85",
    number = "4",
    pages = "437",
    year = "2025"
}

@article{Branco:2011iw,
    author = "Branco, G. C. and Ferreira, P. M. and Lavoura, L. and Rebelo, M. N. and Sher, Marc and Silva, Joao P.",
    title = "{Theory and phenomenology of two-Higgs-doublet models}",
    eprint = "1106.0034",
    archivePrefix = "arXiv",
    primaryClass = "hep-ph",
    doi = "10.1016/j.physrep.2012.02.002",
    journal = "Phys. Rept.",
    volume = "516",
    pages = "1--102",
    year = "2012"
}

@article{ATLAS:2025mim,
    collaboration = "ATLAS",
    title = "{Search for a resonance decaying into a scalar particle and a Higgs boson in the final state with two bottom quarks and two photons with 199 $\fb^{-1}$ of data collected at {\ensuremath{\sqrt{}}}s=13 TeV and {\ensuremath{\sqrt{}}}s=13.6 TeV with the ATLAS detector.}",
    reportNumber = "ATLAS-CONF-2025-009",
    note = "ATLAS-CONF-2025-0090",
    year = "2025"
}

@article{Kauer:2019qei,
    author = {Kauer, Nikolas and Lind, Alexander and Maierh{\"o}fer, Philipp and Song, Weimin},
    title = "{Higgs interference effects at the one-loop level in the 1-Higgs-Singlet extension of the Standard Model}",
    eprint = "1905.03296",
    archivePrefix = "arXiv",
    primaryClass = "hep-ph",
    reportNumber = "FR-PHENO-2019-004",
    doi = "10.1007/JHEP07(2019)108",
    journal = "JHEP",
    volume = "07",
    pages = "108",
    year = "2019"
}

@article{Dawson:2015haa,
    author = "Dawson, S. and Lewis, I. M.",
    title = "{NLO corrections to double Higgs boson production in the Higgs singlet model}",
    eprint = "1508.05397",
    archivePrefix = "arXiv",
    primaryClass = "hep-ph",
    reportNumber = "SLAC-PUB-16335",
    doi = "10.1103/PhysRevD.92.094023",
    journal = "Phys. Rev. D",
    volume = "92",
    number = "9",
    pages = "094023",
    year = "2015"
}

@article{Kauer:2015hia,
    author = "Kauer, Nikolas and O'Brien, Claire",
    title = "{Heavy Higgs signal{\textendash}background interference in $gg\rightarrow VV$ in the Standard Model plus real singlet}",
    eprint = "1502.04113",
    archivePrefix = "arXiv",
    primaryClass = "hep-ph",
    doi = "10.1140/epjc/s10052-015-3586-3",
    journal = "Eur. Phys. J. C",
    volume = "75",
    pages = "374",
    year = "2015"
}

@article{Feuerstake:2024uxs,
    author = "Feuerstake, Finn and Fuchs, Elina and Robens, Tania and Winterbottom, Daniel",
    title = "{Interference effects in resonant di-Higgs production at the LHC in the Higgs singlet extension}",
    eprint = "2409.06651",
    archivePrefix = "arXiv",
    primaryClass = "hep-ph",
    reportNumber = "RBI-ThPhys-2024-15",
    doi = "10.1007/JHEP04(2025)094",
    journal = "JHEP",
    volume = "04",
    pages = "094",
    year = "2025"
}

@article{Ilnicka:2018def,
    author = "Ilnicka, Agnieszka and Robens, Tania and Stefaniak, Tim",
    title = "{Constraining Extended Scalar Sectors at the LHC and beyond}",
    eprint = "1803.03594",
    archivePrefix = "arXiv",
    primaryClass = "hep-ph",
    reportNumber = "DESY 18-031, DESY-18-031",
    doi = "10.1142/S0217732318300070",
    journal = "Mod. Phys. Lett. A",
    volume = "33",
    number = "10n11",
    pages = "1830007",
    year = "2018"
}

@article{Robens:2016xkb,
    author = "Robens, Tania and Stefaniak, Tim",
    title = "{LHC Benchmark Scenarios for the Real Higgs Singlet Extension of the Standard Model}",
    eprint = "1601.07880",
    archivePrefix = "arXiv",
    primaryClass = "hep-ph",
    reportNumber = "SCIPP-16-03",
    doi = "10.1140/epjc/s10052-016-4115-8",
    journal = "Eur. Phys. J. C",
    volume = "76",
    number = "5",
    pages = "268",
    year = "2016"
}

@article{Pruna:2013bma,
    author = "Pruna, Giovanni Marco and Robens, Tania",
    title = "{Higgs singlet extension parameter space in the light of the LHC discovery}",
    eprint = "1303.1150",
    archivePrefix = "arXiv",
    primaryClass = "hep-ph",
    reportNumber = "PSI-PR-13-03",
    doi = "10.1103/PhysRevD.88.115012",
    journal = "Phys. Rev. D",
    volume = "88",
    number = "11",
    pages = "115012",
    year = "2013"
}

@article{Kauer:2012hd,
    author = "Kauer, Nikolas and Passarino, Giampiero",
    title = "{Inadequacy of zero-width approximation for a light Higgs boson signal}",
    eprint = "1206.4803",
    archivePrefix = "arXiv",
    primaryClass = "hep-ph",
    doi = "10.1007/JHEP08(2012)116",
    journal = "JHEP",
    volume = "08",
    pages = "116",
    year = "2012"
}

@article{Martin:2013ula,
    author = "Martin, Stephen P.",
    title = "{Interference of Higgs Diphoton Signal and Background in Production with a Jet at the LHC}",
    eprint = "1303.3342",
    archivePrefix = "arXiv",
    primaryClass = "hep-ph",
    reportNumber = "FERMILAB-PUB-13-677-T",
    doi = "10.1103/PhysRevD.88.013004",
    journal = "Phys. Rev. D",
    volume = "88",
    number = "1",
    pages = "013004",
    year = "2013"
}

@article{Martin:2012xc,
    author = "Martin, Stephen P.",
    title = "{Shift in the LHC Higgs Diphoton Mass Peak from Interference with Background}",
    eprint = "1208.1533",
    archivePrefix = "arXiv",
    primaryClass = "hep-ph",
    reportNumber = "FERMILAB-PUB-12-866-T",
    doi = "10.1103/PhysRevD.86.073016",
    journal = "Phys. Rev. D",
    volume = "86",
    pages = "073016",
    year = "2012"
}

@article{Coradeschi:2015tna,
    author = {Coradeschi, F. and de Florian, D. and Dixon, L. J. and Fidanza, N. and H{\"o}che, S. and Ita, H. and Li, Y. and Mazzitelli, J.},
    title = "{Interference effects in the $H(\rightarrow \gamma\gamma) + 2$ jets channel at the LHC}",
    eprint = "1504.05215",
    archivePrefix = "arXiv",
    primaryClass = "hep-ph",
    reportNumber = "SLAC-PUB-16256, CALT-TH-2015-018, FR-PHENO-2015-003",
    doi = "10.1103/PhysRevD.92.013004",
    journal = "Phys. Rev. D",
    volume = "92",
    number = "1",
    pages = "013004",
    year = "2015"
}

@article{Dixon:2003yb,
    author = "Dixon, Lance J. and Siu, M. Stewart",
    title = "{Resonance continuum interference in the diphoton Higgs signal at the LHC}",
    eprint = "hep-ph/0302233",
    archivePrefix = "arXiv",
    reportNumber = "SLAC-PUB-9654",
    doi = "10.1103/PhysRevLett.90.252001",
    journal = "Phys. Rev. Lett.",
    volume = "90",
    pages = "252001",
    year = "2003"
}

@book{Peskin:1995ev,
    author = "Peskin, Michael E. and Schroeder, Daniel V.",
    title = "{An Introduction to quantum field theory}",
    doi = "10.1201/9780429503559",
    isbn = "978-0-201-50397-5, 978-0-429-50355-9, 978-0-429-49417-8",
    publisher = "Addison-Wesley",
    address = "Reading, USA",
    year = "1995"
}

@article{Buccioni:2019sur,
    author = {Buccioni, Federico and Lang, Jean-Nicolas and Lindert, Jonas M. and Maierh{\"o}fer, Philipp and Pozzorini, Stefano and Zhang, Hantian and Zoller, Max F.},
    title = "{OpenLoops 2}",
    eprint = "1907.13071",
    archivePrefix = "arXiv",
    primaryClass = "hep-ph",
    reportNumber = "IPPP/19/62, FR-PHENO-2019-12, PSI-PR-19-15, ZU-TH 37/19",
    doi = "10.1140/epjc/s10052-019-7306-2",
    journal = "Eur. Phys. J. C",
    volume = "79",
    number = "10",
    pages = "866",
    year = "2019"
}

@article{Baglio:2020wgt,
    author = {Baglio, J. and Campanario, F. and Glaus, S. and M{\"u}hlleitner, M. and Ronca, J. and Spira, M.},
    title = "{$gg\to HH$ : Combined uncertainties}",
    eprint = "2008.11626",
    archivePrefix = "arXiv",
    primaryClass = "hep-ph",
    reportNumber = "CERN-TH-2020-139, IFIC/20-42, FTUV-20-0823, KA-TP-11-2020,
  PSI-PR-20-13, KA-TP-11-2020, PSI-PR-20-13",
    doi = "10.1103/PhysRevD.103.056002",
    journal = "Phys. Rev. D",
    volume = "103",
    number = "5",
    pages = "056002",
    year = "2021"
}

@article{Sherpa:2024mfk,
    author = "Bothmann, Enrico and others",
    collaboration = "Sherpa",
    title = "{Event generation with Sherpa 3}",
    eprint = "2410.22148",
    archivePrefix = "arXiv",
    primaryClass = "hep-ph",
    reportNumber = "IPPP/24/67, LTH-1385, FERMILAB-PUB-24-0748-T, ZU-TH 51/24, MCNET-24-17, CERN-TH-2024-171",
    doi = "10.1007/JHEP12(2024)156",
    journal = "JHEP",
    volume = "12",
    pages = "156",
    year = "2024"
}

@article{Davidson:2005cw,
    author = "Davidson, Sacha and Haber, Howard E.",
    title = "{Basis-independent methods for the two-Higgs-doublet model}",
    eprint = "hep-ph/0504050",
    archivePrefix = "arXiv",
    reportNumber = "IPPP-03-23, DCPT-03-46, SCIPP-04-15",
    doi = "10.1103/PhysRevD.72.099902",
    journal = "Phys. Rev. D",
    volume = "72",
    pages = "035004",
    year = "2005",
    note = "[Erratum: Phys.Rev.D 72, 099902 (2005)]"
}

@inproceedings{Robens:2025nev,
    author = "Robens, Tania and Santos, Rui",
    title = "{BSM: Extended Scalar Sectors}",
    eprint = "2507.21910",
    archivePrefix = "arXiv",
    primaryClass = "hep-ph",
    reportNumber = "RBI-ThPhys-2025-031, CERN-TH-2025-148",
    month = "7",
    year = "2025"
}

@article{Grazzini:2018bsd,
    author = "Grazzini, Massimiliano and Heinrich, Gudrun and Jones, Stephen and Kallweit, Stefan and Kerner, Matthias and Lindert, Jonas M. and Mazzitelli, Javier",
    title = "{Higgs boson pair production at NNLO with top quark mass effects}",
    eprint = "1803.02463",
    archivePrefix = "arXiv",
    primaryClass = "hep-ph",
    reportNumber = "CERN-TH-2018-044, IPPP/18/15, MPP-2018-30, ZU-TH 10/18, IPPP-18-15, ZU-TH-10-18",
    doi = "10.1007/JHEP05(2018)059",
    journal = "JHEP",
    volume = "05",
    pages = "059",
    year = "2018"
}

@article{Robens:2015gla,
    author = "Robens, Tania and Stefaniak, Tim",
    title = "{Status of the Higgs Singlet Extension of the Standard Model after LHC Run 1}",
    eprint = "1501.02234",
    archivePrefix = "arXiv",
    primaryClass = "hep-ph",
    reportNumber = "SCIPP-15-02",
    doi = "10.1140/epjc/s10052-015-3323-y",
    journal = "Eur. Phys. J. C",
    volume = "75",
    pages = "104",
    year = "2015"
}

@article{Carena:2018vpt,
    author = "Carena, Marcela and Liu, Zhen and Riembau, Marc",
    title = "{Probing the electroweak phase transition via enhanced di-Higgs boson production}",
    eprint = "1801.00794",
    archivePrefix = "arXiv",
    primaryClass = "hep-ph",
    reportNumber = "FERMILAB-PUB-17-600-T",
    doi = "10.1103/PhysRevD.97.095032",
    journal = "Phys. Rev. D",
    volume = "97",
    number = "9",
    pages = "095032",
    year = "2018"
}

@article{Denner:2006ic,
    author = "Denner, Ansgar and Dittmaier, S.",
    editor = "Blumlein, J. and Moch, S. and Riemann, T.",
    title = "{The Complex-mass scheme for perturbative calculations with unstable particles}",
    eprint = "hep-ph/0605312",
    archivePrefix = "arXiv",
    reportNumber = "MPP-2006-67, PSI-PR-06-08",
    doi = "10.1016/j.nuclphysbps.2006.09.025",
    journal = "Nucl. Phys. B Proc. Suppl.",
    volume = "160",
    pages = "22--26",
    year = "2006"
}

@article{Denner:2005fg,
    author = "Denner, Ansgar and Dittmaier, S. and Roth, M. and Wieders, L. H.",
    title = "{Electroweak corrections to charged-current e+ e- $\to$ 4 fermion processes: Technical details and further results}",
    eprint = "hep-ph/0505042",
    archivePrefix = "arXiv",
    reportNumber = "MPP-2005-23, PSI-PR-05-05",
    doi = "10.1016/j.nuclphysb.2011.09.001",
    journal = "Nucl. Phys. B",
    volume = "724",
    pages = "247--294",
    year = "2005",
    note = "[Erratum: Nucl.Phys.B 854, 504--507 (2012)]"
}

@article{Denner:1999gp,
    author = "Denner, Ansgar and Dittmaier, S. and Roth, M. and Wackeroth, D.",
    title = "{Predictions for all processes e+ e- $\to$ 4 fermions + gamma}",
    eprint = "hep-ph/9904472",
    archivePrefix = "arXiv",
    reportNumber = "BI-TP-99-10, PSI-PR-99-12",
    doi = "10.1016/S0550-3213(99)00437-X",
    journal = "Nucl. Phys. B",
    volume = "560",
    pages = "33--65",
    year = "1999"
}

@article{Denner:2014zga,
    author = "Denner, Ansgar and Lang, Jean-Nicolas",
    title = "{The Complex-Mass Scheme and Unitarity in perturbative Quantum Field Theory}",
    eprint = "1406.6280",
    archivePrefix = "arXiv",
    primaryClass = "hep-ph",
    doi = "10.1140/epjc/s10052-015-3579-2",
    journal = "Eur. Phys. J. C",
    volume = "75",
    number = "8",
    pages = "377",
    year = "2015"
}

@misc{uhlemann_dipl,
    author = "Uhlemann, Christoph",
    title = "{Narrow-width approximation in the Minimal Supersymmetric Standard Model}",
    year = "2007",
    type = "{Diploma Thesis}",
    howpublished = {\href{https://www.physik.uni-wuerzburg.de/fileadmin/11030200/forschung/publikationen\\/Diplomarbeiten/Uhlemann-dipl.pdf}{Diploma thesis, Wuerzburg}}
   }

@article{Berdine:2007uv,
    author = "Berdine, D. and Kauer, N. and Rainwater, D.",
    title = "{Breakdown of the Narrow Width Approximation for New Physics}",
    eprint = "hep-ph/0703058",
    archivePrefix = "arXiv",
    doi = "10.1103/PhysRevLett.99.111601",
    journal = "Phys. Rev. Lett.",
    volume = "99",
    pages = "111601",
    year = "2007"
}

@article{Gunion:1990kf,
    author = "Gunion, J. F. and Haber, H. E. and Wudka, J.",
    title = "{Sum rules for Higgs bosons}",
    reportNumber = "NSF-ITP-90-127, UCD-90-18, SCIPP-90-15",
    doi = "10.1103/PhysRevD.43.904",
    journal = "Phys. Rev. D",
    volume = "43",
    pages = "904--912",
    year = "1991"
}

@article{Denner:2019vbn,
    author = "Denner, Ansgar and Dittmaier, Stefan",
    title = "{Electroweak Radiative Corrections for Collider Physics}",
    eprint = "1912.06823",
    archivePrefix = "arXiv",
    primaryClass = "hep-ph",
    reportNumber = "FR-PHENO-019",
    doi = "10.1016/j.physrep.2020.04.001",
    journal = "Phys. Rept.",
    volume = "864",
    pages = "1--163",
    year = "2020"
}

@book{Itzykson:1980rh,
    author = "Itzykson, C. and Zuber, J. B.",
    title = "{Quantum Field Theory}",
    isbn = "978-0-486-44568-7",
    publisher = "McGraw-Hill",
    address = "New York",
    series = "International Series In Pure and Applied Physics",
    year = "1980"
}

@book{pilkuhn1967interactions,
  title={The Interactions of Hadrons},
  author={Pilkuhn, H.M.},
  lccn={67026463},
  url={https://books.google.hr/books?id=n2FeAAAAIAAJ},
  year={1967},
  publisher={North-Holland Publishing Company}
}

@article{Dicus:1984fu,
    author = "Dicus, Duane A. and Sudarshan, E. C. G. and Tata, Xerxes",
    title = "{Factorization Theorem for Decaying Spinning Particles}",
    reportNumber = "DOE-ER-03992-567",
    doi = "10.1016/0370-2693(85)91571-0",
    journal = "Phys. Lett. B",
    volume = "154",
    pages = "79--85",
    year = "1985"
}

@inproceedings{Hoang:2024oeq,
    author = {Hoang, Andr\'e H. and Pl\"atzer, Simon and Regner, Christoph and Ruffa, Ines},
    title = "{Beyond the Narrow-Width Limit for Off-Shell and Boosted Differential Top Quark Decays}",
    booktitle = "{16th International Workshop on Top Quark Physics}",
    eprint = "2401.05035",
    archivePrefix = "arXiv",
    primaryClass = "hep-ph",
    reportNumber = "UWThPh-2023-31",
    month = "1",
    year = "2024"
}

@article{Uhlemann:2008pm,
    author = "Uhlemann, C. F. and Kauer, N.",
    title = "{Narrow-width approximation accuracy}",
    eprint = "0807.4112",
    archivePrefix = "arXiv",
    primaryClass = "hep-ph",
    doi = "10.1016/j.nuclphysb.2009.01.022",
    journal = "Nucl. Phys. B",
    volume = "814",
    pages = "195--211",
    year = "2009"
}

@article{Cacciapaglia:2009ic,
    author = "Cacciapaglia, Giacomo and Deandrea, Aldo and De Curtis, Stefania",
    title = "{Nearby resonances beyond the Breit-Wigner approximation}",
    eprint = "0906.3417",
    archivePrefix = "arXiv",
    primaryClass = "hep-ph",
    doi = "10.1016/j.physletb.2009.10.090",
    journal = "Phys. Lett. B",
    volume = "682",
    pages = "43--49",
    year = "2009"
}

@article{Fuchs:2017wkq,
    author = "Fuchs, Elina and Weiglein, Georg",
    title = "{Impact of CP-violating interference effects on MSSM Higgs searches}",
    eprint = "1705.05757",
    archivePrefix = "arXiv",
    primaryClass = "hep-ph",
    reportNumber = "DESY-17-022",
    doi = "10.1140/epjc/s10052-018-5543-4",
    journal = "Eur. Phys. J. C",
    volume = "78",
    number = "2",
    pages = "87",
    year = "2018"
}

@article{Bagnaschi:2018ofa,
    author = "Bagnaschi, Emanuele and others",
    title = "{MSSM Higgs Boson Searches at the LHC: Benchmark Scenarios for Run 2 and Beyond}",
    eprint = "1808.07542",
    archivePrefix = "arXiv",
    primaryClass = "hep-ph",
    reportNumber = "MPP-2018-211, DESY 18-140, DESY-18-140, KA-TP-25-2018, IFT-UAM/CSIC-18-017,
  EFI-18-12, PSI-PR-19-13, IFT-UAM/CSIC-18-017, EFI-18-12",
    doi = "10.1140/epjc/s10052-019-7114-8",
    journal = "Eur. Phys. J. C",
    volume = "79",
    number = "7",
    pages = "617",
    year = "2019"
}

@misc{Fuchs:2015jwa,
    author = "Fuchs, Elina",
    title = "{Interference effects in new physics processes at the LHC}",
    howpublished = "\href{https://inspirehep.net/files/7576e3f3bffbc9e9d528ba5e60fa4aa5}{PhD Thesis, U. Hamburg 2015, DESY-THESIS-2015-037}",
    year = "2015"
}
\end{document}